\documentclass[amsmath,amssymb,aps,twocolumn,prl,floatfix]{revtex4}

\usepackage{mathrsfs}
\usepackage{amsmath}
\usepackage{amssymb}
\usepackage{color}
\usepackage{graphicx}	% Include figure files
\usepackage{bm}% bold math
\usepackage{ulem} % texte barré
\usepackage[squaren, Gray, cdot]{SIunits}
\usepackage{braket}
\usepackage[printwatermark]{xwatermark}
\usepackage{xcolor}
\newcommand{\be}{\begin{equation}}
\newcommand{\ee}{\end{equation}}
\newcommand{\bef}{\begin{figure}}
\newcommand{\eef}{\end{figure}}
\newcommand{\bea}{\begin{eqnarray}}
\newcommand{\eea}{\end{eqnarray}}

\begin{document}
\title{Emergent Properties of Antiagglomerant Films Control Methane Transport: Implications for Hydrate Management}
\author{Fran\c cois Sicard$^{1}$}
\thanks{Corresponding author: \texttt{francois.sicard@free.fr}.}
\author{Tai Bui$^{1}$}
\author{Deepak Monteiro$^{2}$}
\author{Qiang Lan$^{2}$}
\author{Mark Ceglio$^{2}$}
\author{Charlotte Burress$^{2}$}
\author{Alberto Striolo$^{1}$}
\affiliation{$^1$ Department of Chemical Engineering, University College London, WC1E 7JE, London, UK}
\affiliation{$^2$ Halliburton, Houston, Texas, USA}

%\keywords{antiagglomerant, gas hydrate management, flow assurance, numerical simulations, free-energy landscape, transition rates}
%\date{\today}
%	     
\begin{abstract}
The relation between collective properties and performance of antiagglomerants (AAs) used 
in hydrate management is handled using molecular dynamics simulations and enhanced sampling techniques.
A thin film of AAs adsorbed at the interface between one flat sII methane hydrate substrate 
and a fluid hydrocarbon mixture containing methane and $n$-dodecane is studied. 
The AA considered is a surface-active compound with a complex hydrophilic head 
that contains both amide and tertiary ammonium cation groups and hydrophobic tails. 
At sufficiently high AA density, the interplay between the surfactant layer 
and the liquid hydrocarbon excludes methane from the interfacial region. In this scenario,
we combine metadynamics and umbrella sampling frameworks to study accurately the free-energy landscape 
and the equilibrium rates associated with the transport of one methane molecule
 across the AA film. We observe that local configurational changes of the liquid 
hydrocarbon packed within the AA film are associated with high free-energy barriers 
for methane transport. The time scales estimated for the transport of methane across the AA 
film can be, in some cases, comparable to those reported in the literature for the growth of the hydrates, 
suggesting that one possible mechanism by which AAs delay the formation of hydrate plugs 
could be providing a barrier to methane transport. Considering the interplay between structural design 
and collective properties of AAs might be of relevance to improve their performance 
in flow assurance.
\end{abstract}

\maketitle
 
\section{Introduction}
Gas hydrates, also known as clathrate hydrates, are ice-like inclusion compounds consisting of polyhedral 
hydrogen-bonded water cages stabilized by guest gas molecules~\cite{2002-FPE-Koh-Soper,2006-EF-Kelland,2008-CRCPress-Sloan-Koh}. 
They are not chemical compounds because no strong chemical bonds exist between  water 
and gas molecules~\cite{2009-Science-Walsh-Wu,2015-JCP-Michalis-Economou}. They are formed under high-pressure 
and low-temperature conditions such as those found in deep oceans and pipelines~\cite{1997-Geology-Brewer-Kirkwood}. 
The gas molecules able to be trapped (enclathrated) into the water cages are usually small $(<10~\angstrom)$: 
methane, ethane, propane, 1-butane, nitrogen, hydrogen, and carbon dioxide~\cite{2009-IECR-Sum-Sloan}. 
Clathrate hydrates are relevant in a variety of scientific and industrial contexts, 
including climate change modeling~\cite{1996-Paleoceanography-Kaiho-Wada}, 
carbon dioxide sequestration~\cite{2006-PNAS-Park-Yaghi}, hydrocarbon extraction~\cite{2003-Nature-Sloan}, 
hydrogen and natural gas storage~\cite{2003-Nature-Sloan,2004-Science-Florusse-Sloan,2002-Science-Mao-Zhao}, 
separation and refrigeration technologies~\cite{2006-ATE-Ogawa-Mori}, 
marine biology~\cite{2000-Naturwissenschaften-Fisher-McMullin}, and planetary surface chemistry~\cite{1974-Science-Milton}. 
Of particular interest are the hydrocarbon hydrates that can form blockages in oil 
and gas pipelines~\cite{2008-CRCPress-Sloan-Koh,2011-Elsevier-Sloan-Talley}. 
This phenomenon severely affects the safety of pipeline flow assurance~\cite{2011-Elsevier-Sloan-Talley}.\\

Three major stages are associated with hydrate plug formation: 
nucleation~\cite{2008-CRCPress-Sloan-Koh,2001-OxfordPress-Mullin}, growth~\cite{2008-CRCPress-Sloan-Koh,2001-FPE-Freer-Sloan}, 
and agglomeration~\cite{2005-PhD-Turner,2000-RichardsonPress-Sloan}. 
Numerous experimental investigations~\cite{2008-CRCPress-Sloan-Koh,2015-ET-Lee-Lee,2014-Fuel-Fandino-Ruffine,
2013-CGD-Kulkarni-terHorst,2011-EF-Abay-Svartaas,2014-SR-Li-Sum,2016-EF-Ke-Kelland}, 
modeling, and simulations~\cite{2015-FD-Yuhara-Sum,2007-FD-Moon-Rodger,2006-JPCB-Vatamanu-Kusalik,2012-EF-Sum-Sloan,
2008-CES-Ribeiro-Large} have contributed to the current understanding of such stages. 
As offshore drilling activities have moved towards geological sites of deeper waters 
and colder temperatures~\cite{2014-PED-Zhang-Fang}, the community is facing ever-more severe 
technical challenges.
To manage hydrates in pipelines, hydrate inhibitors are used. They are differentiated depending on 
their mode of action: thermodynamic hydrate inhibitors (TIs)~\cite{2003-Nature-Sloan}, 
such as methanol and monoethylene glycol, shift the stability conditions of hydrates 
to lower temperatures and higher pressures, but require large amounts ($10$ to $50$ wt$\%$) 
to be effective.
Low dosage hydrate inhibitors (LDHIs)~\cite{2006-EF-Kelland} instead are effective at concentrations 
as low as $1$ wt$\%$ of water. LDHIs were introduced in the mid-1980s and early 1990s~\cite{1996-CES-Lederhos-Sloan}. 
Unlike TIs, LDHIs do not influence the thermodynamics of hydrates formation but affect its kinetics~\cite{2006-EF-Kelland}. 
Kinetic hydrate inhibitors (KHIs) and antiagglomerants (AAs) are the two main LDHIs classes~\cite{2016-Fuel-Zhao-Firoozabadi}.
Most KHIs are polymeric compounds containing amide groups, such as poly($N$-vinylpyrrolidone),
polyvinyl-caprolactam, and polydiethylacrylamide~\cite{2006-EF-Kelland,1995-Patent-Sloan}. 
They are believed to delay hydrate nucleation and/or growth. 
AAs, mostly surface-active surfactants, are usually amphiphilic chemicals with complex hydrophobic tails 
and hydrophilic headgroups~\cite{2017-Langmuir-Bui-Striolo}. They allow the hydrate particles to form but keep them dispersed, 
yielding transportable slurries~\cite{2011-Elsevier-Sloan-Talley,2006-EF-Kelland,
2009-JPSE-Kelland-Andersen}. When AAs adsorb at the oil-hydrate interface, 
the hydrophobic tails preferably point toward the hydrocarbon phase, possibly inducing an effective 
repulsion when two hydrates approach each other~\cite{2008-CRCPress-Sloan-Koh,2014-PCCP-Aman-Koh,2016-PCCP-Phan-Striolo}. 
When the AAs polar headgroups are adsorbed on the hydrate surface, they could interfere with the hydrate 
growth~\cite{2009-JPSE-Kelland-Andersen,2017-preprint-Bui-Striolo}. 
Quaternary ammonium salts, first developed by Shell in the early 1990s, are the most well-known 
AAs~\cite{2005-EF-Zanota-Graciaa}. Promising performance 
of commercial additives and new AAs have been reported by several research groups~\cite{2013-EF-Chen-etal,
2006-CES-Kelland-Chosa,2009-JPSE-Kelland-Andersen,2012-EF-Sun-Firoozabadi,2014-ECM-Chen-etal,2015-EF-Chen-etal}. 
Most AAs are only effective at low water content (e.g., less than 30$\%$), but some can be effective at high 
water content (up to 80$\%$), such as those reported by Gao~\cite{2009-EF-Gao}.
While the use of AAs is increasing in subsea projects across the industry~\cite{2014-CRCPress-Kelland}, 
their mechanisms of action remain poorly understood. Such understanding 
is necessary to improve their cost effectiveness and expand the range of conditions 
over which their use is safe and convenient.\\

Because classical molecular dynamics (MD) simulations can follow the trajectories of individual molecules, 
MD has been the preferred technique to investigate the formation of hydrates with and without the presence 
of KHIs~\cite{2009-Science-Walsh-Wu,2014-JPCC-Lauricella-Ciccotti,2015-JCP-Lauricella-Ciccotti,
2015-JCP-Michalis-Economou,2016-PNAS-Hall-Kusalik,2016-PCCP-Phan-Striolo,2017-Langmuir-Bui-Striolo}. 
Recent numerical studies have concentrated on the relation between structure and performance of model AAs,
with the emergent molecular-level characterization of the surface adsorption mechanisms of surfactants to hydrates  
considered as a signature of \textit{microscopic} performance~\cite{ICGH9-1,ICGH9-2}.
The coalescence mechanisms of gas hydrate crystals and water droplet have also been studied~\cite{2016-PCCP-Phan-Striolo}. 
Recently, Bui et al.~\cite{2017-Langmuir-Bui-Striolo} related the \textit{macroscopic} performance 
of a class of AAs in flow-assurance applications to the molecular-level properties of the surfactant 
interfacial film. 
Those simulations, compared to experiments, suggested that effective AAs could provide energy barriers in methane transport.\\

In the present work, the authors quantify such energy barriers as experienced by one methane molecule 
diffusing from the hydrocarbon phase to the growing hydrate.
Building on the work of Bui et al.~\cite{2017-Langmuir-Bui-Striolo}, we consider 
the AA that is most effective at excluding methane from the film of AAs formed at 
the water-hydrocarbon interface. This AA was determined to have good performance in laboratory tests 
designed to screen AAs for flow assurance applications.
The metadynamics~\cite{2002-PNAS-Laio-Parrinello} (metaD) and umbrella sampling~\cite{1977-JCP-Torrie-Valleau} (US)
frameworks are combined to study accurately the free-energy (FE) landscape 
and the equilibrium rates associated with the transport mechanisms of one \textit{free} methane molecule
across a densely packed interfacial layer. At sufficiently high AA density, 
we show that the FE barrier is caused by local configurational changes of the liquid hydrocarbon molecules packed 
within the AA film.

\section{Results and Discussion}
\subsection{Interfacial structure}
\begin{figure}[t]
\includegraphics[width=0.8 \columnwidth, angle=-0]{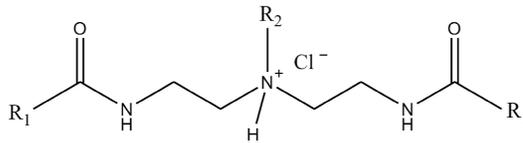}
 \caption{
 Molecular structure of the AA with two long tails $R_1$ ($n$-dodecyl) and one short tail $R_2$ ($n$-butyl). 
 %Reproduced from \cite{2017-Langmuir-Bui-Striolo}.
}
\label{fig1}
\end{figure}
\begin{figure}[b]
\includegraphics[width=0.9 \columnwidth, angle=-0]{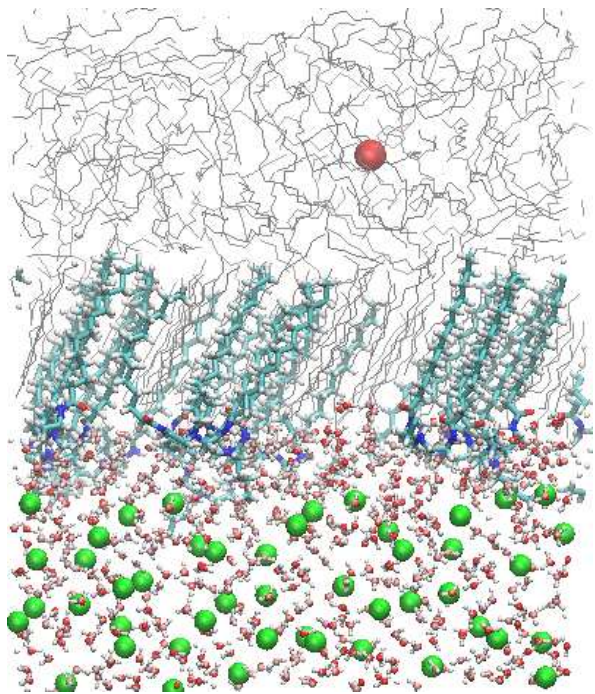}
 \caption{
Initial configuration obtained after equilibration for an AA surface density $\approx 0.67$ molecule/nm$^2$.
Green spheres represent methane molecules in the sII methane hydrate. 
Grey lines represent $n$-dodecane molecules, either in the bulk or trapped within the AA layer. The \textit{free} 
methane molecule is shown as a red sphere. Yellow, red, blue, white, and cyan spheres represent chloride ions, 
oxygen, nitrogen, hydrogen, and carbon atoms in AA molecules, respectively. 
Only half of the simulation box is shown here for clarity.
}
\label{fig2}
\end{figure}

The molecular structure of the AA considered in this study is shown in Fig.~\ref{fig1}, with $R_1$ and $R_2$ 
groups representing long and short tails in one AA, respectively. The AA molecule headgroup includes both 
amide and tertiary ammonium cation groups. Following the notation 
of Bui et al.~\cite{2017-Langmuir-Bui-Striolo}, the AA are denoted as $S_{X}L_{Y}$, 
where $X=4$ ($n$-butyl) and $Y=12$ ($n$-dodecyl) indicate the number of carbon atoms 
in the short (S) and long (L) tail, respectively.
The long AA tails become ordered, almost perpendicular to the hydrate surface, when the surface coverage 
increases, as previously reported by Bui et al.~\cite{2017-Langmuir-Bui-Striolo}.
We consider a surface density $\approx 0.67$ molecule/nm$^2$, sufficiently high to yield 
a dense ordered film at the hydrate-hydrocarbon interface and sufficiently low to allow interface regions 
with both low and high AA density.
To define \textit{unambiguously} the spatial organization of the AA layer, the position of 
the central Nitrogen atom on each surfactant in Fig.~\ref{fig2} was fixed using a restraint potential, 
as implemented in GROMACS~\cite{2015-SoftwareX-Abraham-Lindahl}.
This additional numerical constraint does not impact the analysis of the transport properties 
of the \textit{free} methane molecule through the AA layer as the AAs remain adsorbed 
at the hydrate-hydrocarbon interface within the timeframe of our MD simulations. 
The initial configuration used in the simulations is shown in Fig.~\ref{fig2}.

To characterize the thermodynamic and dynamical properties of the AA film 
adsorbed at the interface between the flat sII methane hydrate and the hydrocarbon fluid, one must 
consider the non-isotropic distribution of the surfactant film at the interface. This spatial distribution 
depends on the AA concentration~\cite{2017-Langmuir-Bui-Striolo} 
and yields  different pathways available to the passage of the \textit{free} methane molecule. 
We quantified the \textit{efficiency} of the AA layer in limiting the transport 
of the methane molecule with the identification of the 
\textit{minimal} FE pathway. This analysis allowed for study of the transport mechanisms at play 
when one \textit{free} methane molecule travels through the interface.

\subsection{Identification of representative pathways}
To identify representative pathways available to the methane molecule  to cross the AA film, 
we ran WT-metaD simulations, as discussed previously. 
This approach was necessary to handle the non-isotropic distribution of the AAs within the interfacial layer. 
However, it did not prevent hysteresis effects caused by \textit{local} reorganization of the liquid hydrocarbons  
within the AA layer. 
This hysteresis effect might occur when the bias potential becomes sufficiently 
high to expel hydrocarbon molecules from the interfacial layer.  
To prevent this possible shortcoming, one should consider an additional CV, 
accounting for the \textit{local} behaviour of the liquid hydrocarbon molecules. 
This is well known, for instance, in biological docking 
studies~\cite{2012-PNAS-Limongelli-Parrinello,2015-PNAS-Tiwary-Berne,2016-PCCP-Yang-Marti,2017-JCAMD-Bhakat-Soderhjelm}.  
Computational limitations made it unfeasible to consider an additional CV 
in the present work.\\ 
\begin{figure}[t]
\includegraphics[width=0.95 \columnwidth, angle=-0]{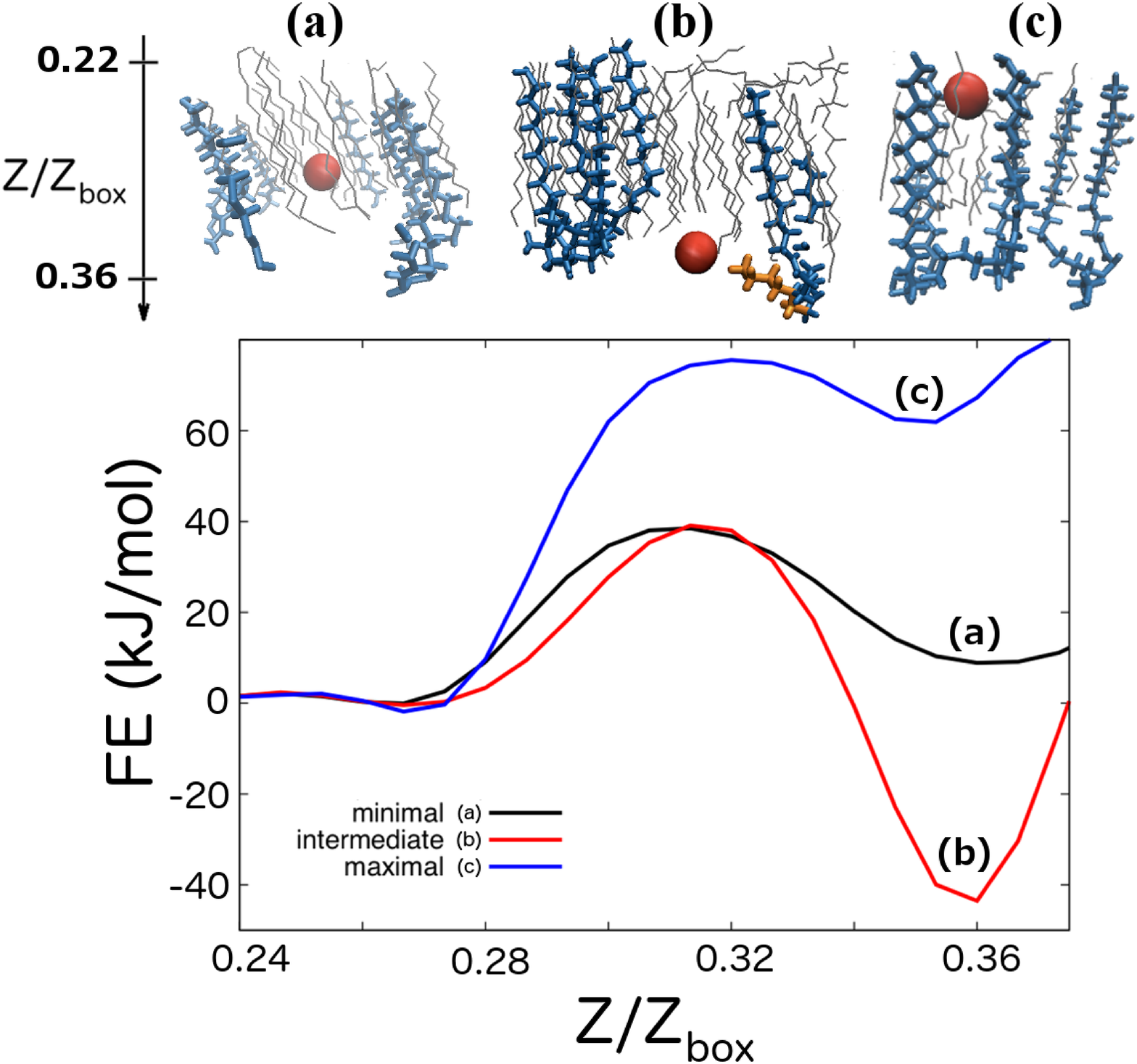}
 \caption{
Bottom: FEPs associated with three pathways obtained with 
the WT-metaD framework. Top: representative snapshots to illustrate the \textit{minimal} (a), 
\textit{intermediate} (b), and \textit{maximal} (c) scenarios. Grey and blue lines represent $n$-dodecane 
and AA molecules, respectively. In the snapshots, the \textit{free} methane molecule is shown as a red sphere.
In (b), the AA short tail interacting with the methane molecule is colored orange. 
The $x$-axis corresponds to the $Z$-Cartesian coordinate of methane 
expressed in reduced units, $Z/Z_{\textrm{box}}$, with $Z_{\textrm{box}}$ the size of the simulation box along the $Z$ direction.
}
\label{fig3}
\end{figure}
\begin{figure}[b]
\includegraphics[width=0.9 \columnwidth, angle=-0]{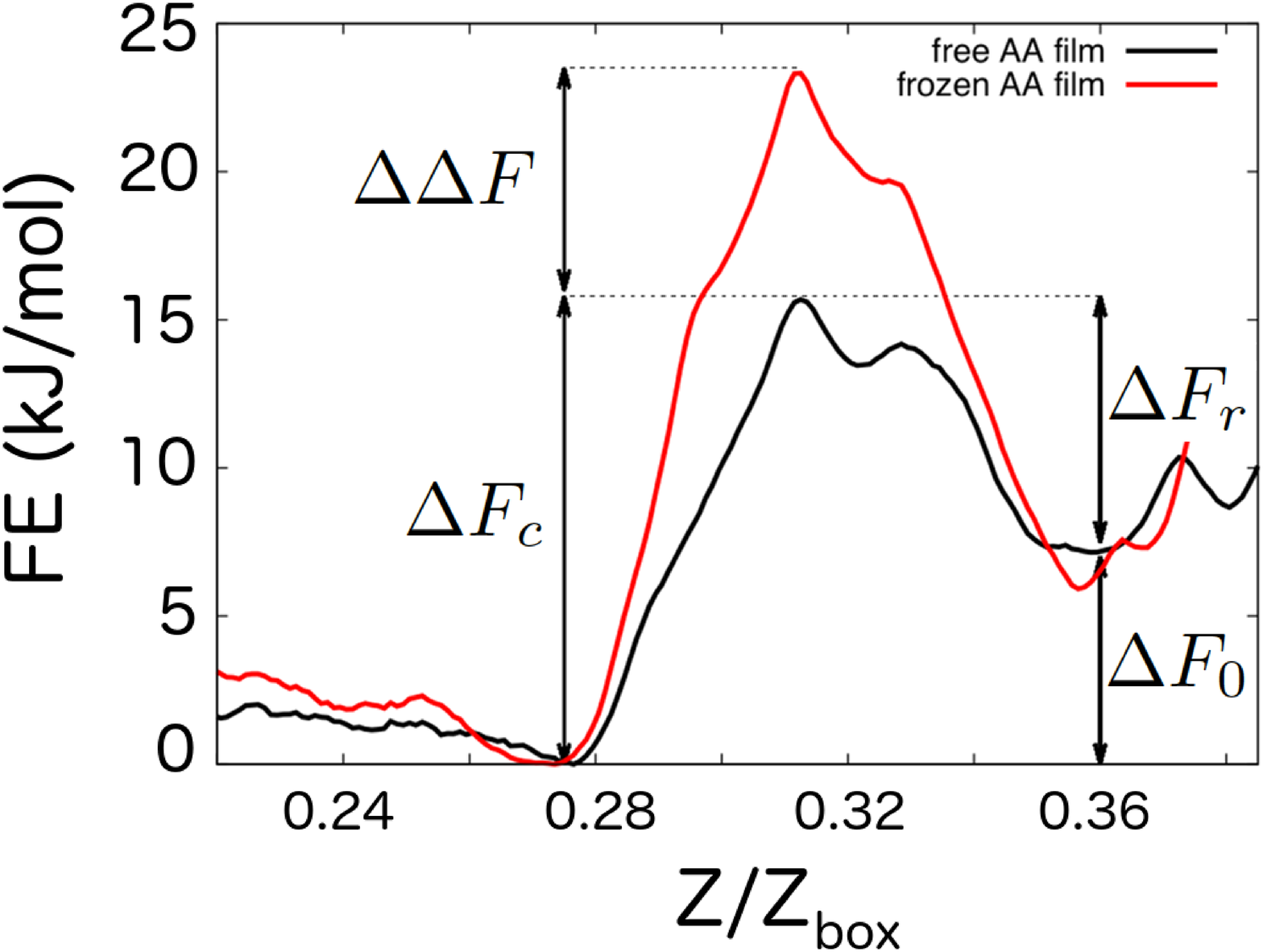}
 \caption{
FEPs associated with the passage of the \textit{free} methane molecule across 
the interfacial layer, obtained within the US/ABMD framework. The AA surface 
density is $\approx 0.67$ molecules/nm$^2$. The $x$-axis corresponds to the $Z$-Cartesian coordinate of methane 
expressed in reduced units, $Z/Z_{\textrm{box}}$, with $Z_{\textrm{box}}$ the size of the simulation box along the $Z$ direction.
We show the \textit{minimal} FE pathways obtained when the AA molecules are either frozen (red) 
or free (black). The activation energies associated with methane capture and escape, $\Delta F_c$ and $\Delta F_r$, 
are $\approx 15.5$ kJ/mol and $\approx 7$ kJ/mol, respectively.
}
\label{fig4}
\end{figure}

Instead, we ran WT-metaD simulations to identify representative pathways. 
Figure~\ref{fig3} shows representative FEPs obtained within 
the WT-metaD framework, along with the respective snapshots to illustrate the different scenarios. 
The pathways identified occur in interfacial regions with different AA density.
The \textit{minimal} FEP corresponds to the diffusion of the methane through an interfacial region 
made up of a large cluster of hydrocarbons. These regions show a characteristic size $\approx 20 \angstrom$ 
in the simulations, which is comparable 
to the molecular dimension of the dodecane molecules (cf. Figure~\ref{fig3}a). 
When hydrocarbons are confined in these narrow regions, it is possible that the effective viscosity 
differs with respect to that in the bulk~\cite{1992-SPRINGER-Granick}. 
The \textit{intermediate} FEP corresponds to the diffusion of the methane along the edge of 
a hydrocarbon cluster to the surrounding AA molecules. 
As shown in Figure~\ref{fig3}b, the methane molecule travels through the interfacial layer 
and eventually interacts with the hydrophobic short tail of one AA, which can be parallel to the interfacial layer. 
This specific orientation enhances the interaction between the short tail and the methane molecule, 
stabilizing the system. 
This effect yields the \textit{inversion} of the thermodynamic stability 
of the system with respect to the global ($Z/Z_{\textrm{box}}\approx 0.27$) 
and local ($Z/Z_{\textrm{box}}\approx 0.36$) minima observed along the \textit{minimal} FEP.
The \textit{maximal} FEP corresponds to the diffusion of
methane through a locally dense AA region, in which a few oil molecules are trapped within the AA layer 
(cf. Figure~\ref{fig3}c). 
Possibly because the AA molecules are rigid in these simulations, the transport of methane across this region 
encounters a large FE barrier. 

As discussed in the Supporting Information (SI), the FEP obtained within the metadynamics framework 
shows the same qualitative behavior as the FEP obtained within the ABMD/US framework, 
with quantitative differences caused by numerical artifacts.
The remainder of the paper discusses the analysis on the \textit{minimal} FE pathway, 
which shows the lowest FE barriers for transport of methane from the hydrocarbon to the 
hydrate (capture) and from the hydrate to the hydrocarbon (escape). In the following detailed analysis, 
we implemented the US and ABMD frameworks, as described previously. 
The \textit{minimal} FEP would provide the lowest boundary for time constants representative for methane transport 
across the AAs film.

\subsection{Minimal free-energy pathway}
\subsubsection{Thermodynamic properties}
Two different systems were compared:  
1) the AA layer is kept frozen, as in the case of the preliminary WT-metaD simulation 
and 2) the AA layer is flexible, although the central nitrogen atom of each AA molecule remains frozen. 
The former approach allows us to define 
\textit{unambiguously} the Cartesian position of the \textit{free} methane molecule with respect to the AA layer. 
\begin{figure*}[t]
\includegraphics[width=0.95 \textwidth, angle=-0]{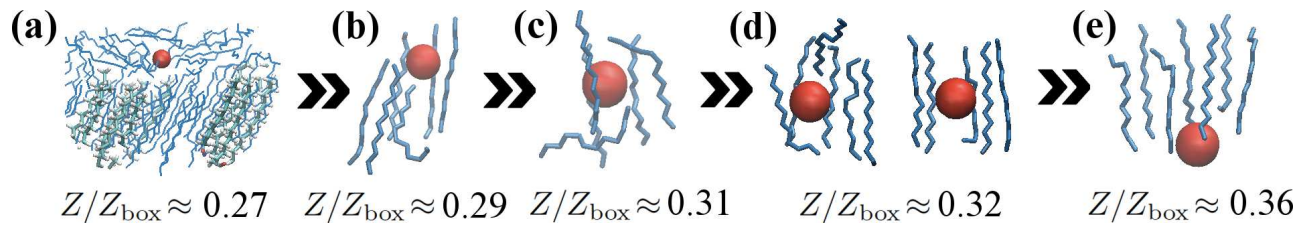}
 \caption{
Sequence of simulation snapshots representing the transport mechanism of the methane molecule (red sphere)
along the \textit{minimal} FE pathway across the flat interfacial layer composed of a mixture of 
AA and dodecane (blue molecules). The bulk hydrocarbon phase and the sII hydrate are 
not shown for clarity. The AA layer is only shown in (a). The methane molecule initially starts 
in the bulk hydrocarbon phase, above the AA layer (a). The methane molecule then enters the interfacial layer 
through two oil molecules (b). As the methane goes farther across the interfacial layer, 
the oil molecules below the methane molecule bend, eventually forming  a \textit{cage} surrounding 
the methane molecule (c). As the methane travels farther down, one oil molecule forming the gate begins 
pushing the methane molecule. At that stage, the system evolves between two states corresponding to a closed 
or opened gate below the methane molecule (d). Eventually, the methane molecule is driven underneath the AA layer (e).
}
\label{fig5}
\end{figure*}
Figure~\ref{fig4} compares the FEP associated to the \textit{minimal} FE path, 
as obtained in the two approaches. In both cases, the AA molecules induce the alignment and rigidity of 
the oil molecules within the interface, as shown in Figure~\ref{fig2} 
and reported 
by Bui et al.~\cite{2017-Langmuir-Bui-Striolo}. Freezing the AA molecules does not impact 
the location of  global and local minima, $Z/Z_{\textrm{box}} = 0.27 $ and $Z/Z_{\textrm{box}}= 0.36$, respectively, 
and the FE $\Delta F_0 \approx 8.5$ kJ/mol. This was expected as these quantities correspond to 
AA-free locations above and below the interfacial layer, which are not impacted by the AA freezing. 
However, freezing the AA molecules yields an increase of the FE barriers, $\Delta \Delta F$ $\equiv \Delta F_c^{\textrm{frozen}} - \Delta F_c^{\textrm{free}} \approx 7.5$ kJ/mol at the transition, 
$Z/Z_{\textrm{box}} \approx 0.31$ (cf. Figure~\ref{fig4}).
This difference highlights the interplay between the rigidity of the AA layer and the \textit{local pliability} 
of the oil molecules trapped within the interfacial layer.

\subsubsection{Transport Mechanisms}
Because the results obtained  when the AA layer is flexible are more realistic, 
we further study the transport mechanism as observed within the US/ABMD framework. 
Figure~\ref{fig5} shows representative snapshots extracted from 
the FEP.
The free methane molecule is initially in the bulk hydrocarbon phase, above the AA layer. 
When it comes closer to the interface, it is first trapped in a local FE minimum ($Z/Z_{\textrm{box}}\approx 0.27$).
This minimum corresponds to a transition region between oil molecules isotropically oriented in the bulk  
and oil molecules parallel to the AA tails. 
The methane molecule then enters the AA layer between two hydrocarbon molecules. As the methane travels farther 
across the interfacial layer, an energy barrier arises as the oil molecules are displaced from the methane pathway.
The transport proceeds until one oil molecule cannot be pushed farther down ($0.27 \lesssim Z/Z_{\textrm{box}} \lesssim 0.31$). 
At these conditions, the oil molecule being displaced bends, eventually forming a \textit{cage} 
surrounding the methane molecule ($Z/Z_{\textrm{box}}\approx 0.31$ ). This corresponds to the 
high energy transition region in the FE landscape.
From this point on, the hydrocarbon forming the \textit{cage} effectively contributes to actively pushing 
the methane towards the hydrate. 
The system evolves between two states corresponding to a closed or opened \textit{gate} below the methane molecule 
($0.31 \lesssim Z/Z_{\textrm{box}} \lesssim 0.34$). 
Once the methane molecule overcomes this \textit{transition state}, it is pushed down underneath 
the AA layer ($Z/Z_{\textrm{box}} \approx 0.35$). The methane molecule then reaches the local 
minimum corresponding to the water layer between the AA layer and the hydrate. 
This analysis does not explore further transport through the water film.

\subsubsection{Dynamical Properties}
To estimate the transition rates along 
the FE pathway, we considered the Kramers theory~\cite{1940-Physica-Kramers,1990-RMP-Hanggi-Borkovec,
2005-Chaos-Pollak-Talkner,2010-QRB-Zhou,2017-ELSEVIER-Peters,2018-arXiv-Sicard}, which is based on the inertial Langevin 
equation~\cite{2012-WS-Coffey-Kalmykov}:
\begin{equation}
m\ddot{q} = -\frac{\partial V}{\partial q} - m \gamma \dot{q} + R(t) \,.
\label{Langevin}
\end{equation}
In Eq.~\ref{Langevin}, $q$ represents the reaction coordinate, $m$ is the reduced mass for the reaction coordinate, 
$\gamma$ is the friction coefficient, and $V(q)$ is a PMF. $R(t)$ is a random force with 
zero mean that satisfies the fluctuation theorem~\cite{2008-PR-Marini-Vulpiani}. 
In the presented case, Eq.~\ref{Langevin} is applied to the methane molecule, the reaction coordinate is the $Z$-Cartesian 
coordinate, $m$ represents the methane molecule mass, 
and $V(q)$ is the PMF discussed in Figure~\ref{fig4}.
The escaping/capture rates can be described within the Kramers theory framework in terms of diffusion across a FE barrier of height 
$\Delta F$ when the barrier (and local minima) are modeled as parabolic potentials~\cite{2017-ELSEVIER-Peters}:
\begin{equation}
V(q) = V_{TS} - \frac{1}{2}m\omega_{TS}^2 (q-q_{TS})^2 \,.
\end{equation}
The Kramers theory provides a physical derivation of the reaction rate constants in terms of the shape of the 
energy profile. 
From intermediate to high friction regimes, the reaction rate, $k$, is given as:
\begin{equation}
\label{Kramers-fullEQ}
k = \frac{\gamma}{\omega_{TS}} \Bigg( \sqrt{\frac{1}{4}+\frac{\omega_{TS}^2}{\gamma^2}}-\frac{1}{2} \Bigg) 
\frac{\omega_0}{2\pi} e^{-\Delta F/k_B T}
\end{equation}
where $\omega_0$ and $\omega_{TS}$ represent the stiffness of the potential well and the barrier, 
respectively, $\gamma=6\pi\eta R/m$ the friction coefficient, with $\eta$ the effective viscosity, 
and $\Delta F$ the FE barrier.
\begin{figure}[b]
\includegraphics[width=0.9 \columnwidth, angle=-0]{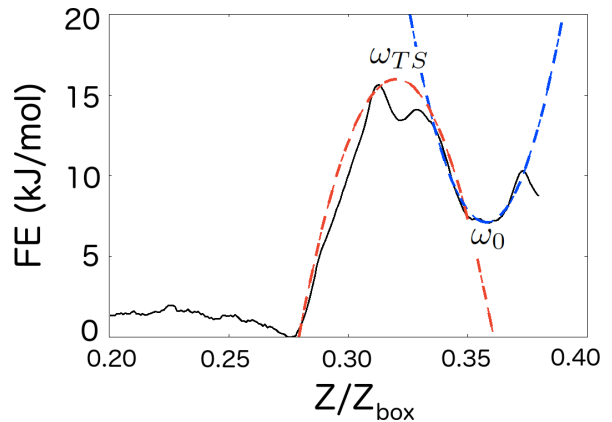}
 \caption{
Fitting of the FEP within parabolic potentials to extract parameters for the Kramers theory. 
The nonlinear least-squares Marquardt-Levenberg 
algorithm was implemented obtaining  
$\omega_0 \approx 100 \times 10^{13}~s^{-1}$ and $\omega_{TS} \approx 230 \times 10^{13}~s^{-1}$.
}
\label{fig6}
\end{figure}
Figure~\ref{fig6} shows the FEP fitted using harmonic potentials 
 at the local minimum and maximum. This yields  
 $\omega_0 \approx 100 \times 10^{13}~s^{-1}$ and $\omega_{TS} \approx 230 \times 10^{13}~s^{-1}$ 
(cf. details in the SI).
One important parameter is the effective viscosity $\eta$. Because of confinement 
of the liquid hydrocarbon at the interface, the effective viscosity is intrinsically different 
from its bulk value~\cite{1992-SPRINGER-Granick} ($\approx 2$ mPa.s at $277$ K and $20$ MPa). 
Singer and Pollock~\cite{1992-SPRINGER-Granick} showed that the effective viscosity can increase 
by several orders of magnitude when dodecane molecules are confined in a space approaching 
the molecular dimension of the liquid hydrocarbon. Eventually, they observed the divergence of the effective viscosity.
Considering $\eta \approx 10^2-10^5$ Pa.s~\cite{1992-SPRINGER-Granick}, 
one obtains a lower boundary for the friction coefficient, 
 $\gamma > 15\times 10^{18}~s^{-1} >> \omega_{TS}$, characteristic of the strong friction regime of interest here.
In this limit, Eq.~\ref{Kramers-fullEQ} simplifies to~\cite{2017-ELSEVIER-Peters}
\begin{equation}
\label{Kramers-shortEQ}
k = \frac{\omega_0 \omega_{TS}}{2\pi \gamma} e^{-\Delta F/k_B T} \,.
\end{equation}
This approximation is often used to interpret, for instance, 
the time scale characteristic for protein or DNA conformational dynamics~\cite{2012-PNAS-Yu-Woodside,2015-JCP-Sicard-Manghi}.
Considering the parameter values discussed previously, one obtains a characteristic time scale for  
methane escape from the aqueous film near the hydrate to the hydrocarbon liquid phase, 
across the AA film, $\tau_{esc} \approx 0.01-1~\mu$s. \\

Building on the recent method of Parrinello, Salvalaglio, and Tiwary~\cite{2013-PRL-Tiwary-Parrinello,
2014-JCTC-Salvalaglio-Parrinello}, we extend the metaD scope to assess numerically $\tau_{esc}$. 
We indicate our numerical result as $\tau_{\textrm{esc}}^{(\textrm{num})}$ in what follows.
WT-metaD was performed using the $Z$ coordinate as CV with the parameters 
reported in the Methods section. The statistics for 
$\tau_{\textrm{esc}}^{(\textrm{num})}$ conformed to a Poisson distribution with mean $\mu=0.08 \pm 0.02~\mu$s 
and variance $\lambda=0.09~\mu$s. The statistics obeys a two-sample KS test with $p$-value equal to $0.52$.
This numerical result confirms the reliability of the choice of the $Z$-Cartesian coordinate 
as a collective variable in the US framework (cf. discussion in the SI) and is comparable to the 
estimation derived from the Kramers theory with a friction coefficient $\gamma \approx 10^4$~Pa.s.

Finally we quantified the characteristic time scale for methane capture
(from the hydrocarbon to the hydrate surface) along the \textit{minimal} FE path, $\tau_{cap}$. 
As shown in Figure~\ref{fig6}, no confinement was present above the AA layer ($Z/Z_{\textrm{box}} < 0.27$). 
Thus the entropic effect would dominate, and the associated rate of capture would tend to infinity. 
To correct for this, an external potential was added at $Z/Z_{\textrm{box}} \approx 0.25$, 
similar to the FE barrier observed for $Z/Z_{\textrm{box}} \in [0.27,0.31]$. 
This external potential would represent, for instance, a free methane molecule confined between 
two AA layers. 
The statistics of our numerical analysis for $\tau_{\textrm{cap}}^{(\textrm{num})}$ conformed to 
a Poisson distribution with mean $\mu=0.35 \pm 0.15~\mu$s and variance $\lambda=0.49~\mu$s, 
and followed a two-sample KS test with $p$-value equal to $0.51$
(cf details in the SI).

\subsection{Intermediate free-energy pathway}

To conclude this analysis, we considered the \textit{intermediate} path defined in Figure~\ref{fig3}, 
which presents a similar FE barrier associated with methane capture but
a deeper energy basin at $Z/Z_{\textrm{box}} \approx 0.35$ along with a higher FE barrier 
associated with methane escape. We reconstructed the accurate FEP within the US/ABMD framework 
and obtained the respective values for $\omega_{TS} \approx 100 \times 10^{13}~s^{-1}$ 
and $\omega_{0} \approx 130 \times 10^{13}~s^{-1}$. The details are reported in the SI. Considering Eq.~\ref{Kramers-shortEQ} 
with $\gamma \approx 10^4$~Pa.s, we obtained a characteristic time scale for methane escape, 
$\tau_{esc} \approx 40~\mu$s. The estimated characteristic time scale for methane capture was similar 
to that obtained in correspondence of the \textit{minimal} path.

%\section{Discussion}
\subsection{Discussion}
The extensive simulations discussed previously allowed for identification of three possible pathways 
that a methane molecule can follow to diffuse from the hydrocarbon phase to the hydrate structure 
across the interfacial film rich in AAs. We focused on the pathway that showed the minimal FE
barrier and we quantified accurately the FEPs using a combination of enhanced sampling techniques. 
This allowed us to reveal the molecular mechanism responsible for the FE barrier. Interpreting the 
FEP within the Kramers theory allowed us to extract the time constant that quantifies 
the escape rate of the methane across the AA film. The result is consistent with estimations from a 
direct metaD estimate, which supports the reliability of our approach.

We showed that the FE barrier comes from local flexibility of the 
liquid hydrocarbon molecules within the AA film.
The flexibility of the oil molecules is impacted by the rigidity of the AA molecules.
We observed that the interaction of the methane molecule with the AA short tail can increase 
the stability of the system and invert its thermodynamic stability.
Increasing the AA surface density leads to an increase in the heigh of the FE barrier, 
improving eventually the efficiency of the interfacial layer in limiting the transport of methane. 

The characteristic time obtained from this study can be compared with the experimental 
growth rate of methane hydrate film at flat methane/water interface. 
Sun et al.~\cite{2010-ARPCSC-Sun-Chen}, and more recently Li et al.~\cite{2014-SR-Li-Sum} 
reported growth characteristic time $\tau^{\textrm{exp}}\approx 1-5$ $\mu$s, depending on the degree of supercooling 
considered in the experiments.
This is one order of magnitude higher than the escaping time obtained in our simulations for 
the \textit{minimal} FE pathway and one order of magnitude lower than that obtained 
for the \textit{intermediate} FE pathway. 
These data suggest that ordering of the long AA tails observed for the AAs considered here can provide effective 
barriers to methane transport. Further, the results suggest that the short tails of the AAs can be designed 
to enhance the stability of methane near the growing hydrate.

The new physical insights discussed in this paper could be useful for a variety of applications. 
For instance, several research groups focus on the interplay between design and performance of AAs 
at the microscopic scale. Accounting for the collective effect of new AAs could allow us to
infer their use in practical applications, which addresses the current needs.

\section{Methods}
\subsection{Unbiased MD simulations}
MD simulations were performed with the GROMACS software 
package, version 5.1.1~\cite{2015-SoftwareX-Abraham-Lindahl} using the TIP4P/Ice water model~\cite{2005-JCP-Abascal-Vega}.
Biased simulations were performed using version 2.3 of the plugin for FE calculation, 
 PLUMED~\cite{2014-CPC-Tribello-Bussi}.
The TIP4P/Ice model has been successfully implemented to study 
hydrate nucleation and growth~\cite{2009-Science-Walsh-Wu,2010-JPCB-Jensen-Sum}  
and to investigate the performance of potential hydrate inhibitors~\cite{2015-PCCP-Alireza-Englezos}. 
This model yields an equilibrium temperature for the formation of gas hydrates at high pressure close to 
experimental values~\cite{2010-JCP-Conde-Vega}. Methane and $n$-dodecane were represented within the united-atom version 
of the TraPPE-UA force field~\cite{1998-JPCB-Martin-Siepmann}. AAs were modeled using the general 
Amber force field (GAFF)~\cite{2004-JCC-Wang-Case}, which is often implemented for modeling organic 
and pharmaceutical molecules containing H, C, N, O, S, P, and halogens. Atomic charges were calculated 
with the AM1-BCC method employed in Antechamber from the Amber 14 suite~\cite{AMBER14}. 
The chloride counterions ($Cl^{-}$) were modeled as charged Lennard-Jones (LJ) spheres with the potential 
parameters taken from Dang~\cite{1994-JCP-Smith-Dang}, without polarizability.
The sII hydrates were considered to be the solid substrate, and they were not allowed to vibrate in this work. 
AAs, chloride counterions, $n$-dodecane, and methane composed the liquid phase. 
Dispersive and electrostatic interactions were modeled by the $12-6$ LJ 
and Coulombic potentials, respectively. The Lorentz-Berthelot mixing rules~\cite{1881-AP-Lorentz,1898-CR-Berthelot} 
were applied to determine the LJ parameters for unlike interactions from the parameters of the pure components. 
The distance cutoff for all non-bonded interactions was set to $1.4$ nm. Long-range corrections 
to the electrostatic interactions were described using the particle mesh Ewald (PME) 
method~\cite{1993-JCP-Darden-Pedersen,1995-JCP-Essmann-Berkowitz,2001-CPL-Kawata-Nagashima} with 
a Fourier grid spacing of $0.12$ nm, a tolerance of $10^{-5}$, and fourth-order interpolation. 
Periodic boundary conditions were applied in three dimensions for all simulations.

To construct the initial configurations, we followed the procedure  described by Bui et al.~\cite{2017-Langmuir-Bui-Striolo}.
One unit cell of sII methane hydrates was adapted from the study of Takeuchi et al.~\cite{2013-JCP-Takeuchi-Yasuoka}. 
The sII methane hydrate unit cell was replicated three times in the $X$ and $Y$ directions ($5.193$ nm) 
and two times in the $Z$ direction ($3.462$ nm). It was then flanked by a thin liquid water film of approximately 0.5 nm 
on both sides along the Z direction. 
The desired number of AA molecules was arranged near both sides of the hydrate substrate. 
The chloride counterions ($Cl^-$) were placed next to the AA headgroups. The $n$-dodecane and methane molecules 
were placed within the remainder of the simulation box.
The time step used in all the simulations was $0.001$ ps, and the list of neighbors was updated every $0.01$ ps 
with the \textit{grid} method and a cutoff radius of $1.4$ nm.

The energy of the model was first optimized with the ``steepest descent minimization" algorithm to remove high-energy 
configurations. 
Subsequently, to minimize the possibility that the initial configuration biased the simulation results, 
an $NVT$ temperature-annealing procedure, as implemented in GROMACS~\cite{2015-SoftwareX-Abraham-Lindahl}, 
was conducted. The algorithm linearly 
decreased the system temperature from $1000$ K to $277$ K in $500$ ps. In these simulations, the hydrate 
substrate and chloride ions were kept frozen. To relax the structure of $n$-dodecane and AAs, 
a $NVT$ simulation was conducted at $277$ K for $2$ ns using the Berendsen 
thermostat~\cite{1984-JCP-Berendsen-Haak}, with the sII hydrate structure kept frozen.
The equilibration phase was then conducted within the isobaric-isothermal ($NPT$) ensemble under thermodynamic 
conditions favorable for hydrate formation ($T=277$ K and $P=20$ MPa) to equilibrate the density.
During the NPT simulation, all molecules in the system were allowed to move, including water and methane molecules in the hydrate substrate.
The pressure coupling was applied only along the $Z$ direction of the simulation box, which allowed  
the $X$ and $Y$ dimensions to be maintained constant. Temperature and pressure were maintained at $277$ K and $20$ MPa, 
respectively using the Berendsen thermostat and barostat~\cite{1984-JCP-Berendsen-Haak} for $5$ ns. 
This is considered the most efficient algorithm to scale simulation boxes at the beginning 
of a simulation~\cite{2013-Bioinformatics-Pronk-Lindakl}. We then switched to 
the Nose-Hoover thermostat~\cite{1985-JCP-Evans-Holian} and the Parrinello-Rahman 
barostat~\cite{1981-JAP-Parrinello-Rahman} for $100$ ns, which are considered 
 more thermodynamically consistent algorithms~\cite{2013-Bioinformatics-Pronk-Lindakl}. 
This numerical protocol allowed the AAs to assemble and orient to form the ordered layer described 
in the work of Bui et al.~\cite{2017-Langmuir-Bui-Striolo}.
The system was then equilibrated for $3$ ns in $NVT$ conditions coupling with the \textit{v-rescale} 
thermostat~\cite{2007-JCP-Bussi-Parrinello} ($T=277$ K, $\tau_T=0.1$ ps). 
To define the position of the \textit{free} methane molecule 
in the simulation box with respect to the sII hydrate structure, the simulation was continued in $NVT$ conditions 
holding in place the methane molecule enclathrated into the water cages defining the sII hydrate structure, 
as implemented with the \textit{freeze group} procedure in GROMACS~\cite{2015-SoftwareX-Abraham-Lindahl}.

\subsection{Biased MD simulations}
The phenomenon of interest (\textit{i.e.} the transport of  methane across the interfacial layer) 
occurs on time scales that are orders of magnitude longer than the accessible time 
that can be currently simulated with classical MD simulations. A variety of methods, referred 
to as \textit{enhanced sampling techniques}~\cite{2006-CR-Adcock-McCammon,2015-BA-Spiwok-Hosek,2015-BBA-Bernardi-Schulten,
2016-Plos-Maximova-Shehu,2017-RP-Pietrucci}, can be implemented to overcome this limitation. These methods 
accelerate rare events and are based on constrained MD. 
MetaD~\cite{2002-PNAS-Laio-Parrinello,2008-RPP-Laio-Gervasio,2011-CMS-Barducci-Parrinello,2012-CMS-Sutto-Gervasio} 
 and US~\cite{1977-JCP-Torrie-Valleau,2011-CMS-Kastner} belong to this class of methods: 
they enhance the sampling 
of the conformational space of a system along a few selected degrees of freedom, named reaction coordinates 
or collective variables (CVs), and reconstruct the probability distribution as a function of these CVs. 
These techniques are proven powerful tools 
to study biological~\cite{2006-JACS-Barducci-Parrinello,2013-PNAS-Sutto-Gervasio,2013-JCP-Sicard-Senet,2015-JCP-Sicard-Manghi} 
and chemical systems~\cite{2015-IUCrJ-Giberti-Parrinello,2015-FD-Salvalaglio-Parrinello,2017-JCP-Gimondi-Salvalaglio}. 
However, despite these successes, 
care should be taken to properly choose and implement the reaction coordinates~\cite{2008-RPP-Laio-Gervasio,2011-CMS-Barducci-Parrinello,2012-CMS-Sutto-Gervasio}.\\

We first ran a well-tempered metaD (WT-metaD) simulation~\cite{2008-PRL-Barducci-Parrinello,
2014-PRL-Dama-Voth} using the three Cartesian coordinates ($X$, $Y$ and $Z$) of the \textit{free} methane molecule as CVs. 
WT-metaD is a method  based on a biasing of the potential surface. The biasing potential is dynamically 
placed on top of the underlying potential energy landscape to discourage the system from visiting the same 
points in the configurational space. The WT-metaD time-dependent bias, $V_{\textrm{bias}}(s,t)$ can have any form, 
but a Gaussian potential is usually implemented~\cite{2008-RPP-Laio-Gervasio}:
\begin{equation}
V_{\textrm{bias}}(s,t) = \omega \sum_{t'<t} \exp\Big[-\frac{(s(t)-s(t'))^2}{2 \sigma^2} \Big]~.
\label{metaD-potential}
\end{equation}
In Eq.~\ref{metaD-potential}, $\omega$ is the height of the biasing potential, $\sigma$ is the width, $t$ is the time, 
and $s$ is the collective 
variable. Following the algorithm introduced by Barducci et al.~\cite{2008-PRL-Barducci-Parrinello}, 
a Gaussian-shaped potential is deposited every $\tau_G = 2$ ps, with height $\omega = \omega_0 e^{-V(s,t)/(f-1)T}$, 
where $\omega_0 = 5$ kJ/mol is the initial height, 
$T$ is the temperature of the simulation, and $f\equiv(T+\Delta T)/T =25$ is the bias factor with $\Delta T$ 
a parameter with the dimension of a temperature. 

In our implementation, the resolution of the recovered FE surface is 
determined by the width of the Gaussian $\sigma=0.25$ nm along the $X$, $Y$ and $Z$ directions.
This step was motivated by the inherent competition between different pathways that can lead to methane 
transport through the AA layer, 
with each path characterized by different FE barriers. 
WT-metaD simulations were run restraining the position of the AA layer while allowing the hydrocarbon molecules, both
trapped at the interface and present in the bulk phase, to move freely. This was necessary to minimize \textit{local} hysteresis 
effects that would occur in the \textit{global} convergence of the WT-metaD resulting from the flexibility 
of the AA layer in defining the Cartesian position of the \textit{free} methane molecule with respect to the AA film.

Once the possible pathways across the AA film were identified, the potential of mean force (PMF) 
along them was rigorously calculated using US. 
This combined metaD/US approach, originally proposed by Zhang and Voth~\cite{2011-JCTC-Zhang-Voth} 
in the biophysical context, provides a powerful means to calculate a  physically meaningful PMF where the convergence 
problem, sometimes associated with metaD, is avoided. The approach also avoids the discontinuity problem, 
often associated with US calculations that assume a straight-line reaction coordinate, 
yielding a  physically accurate PMF~\cite{2011-JCTC-Zhang-Voth}.
It is worth noting that in the combined metaD/US approach, the final PMF is not sensitive to the choice of the WT-metaD 
parameters, as long as they are in reasonable range because the PMF is not directly calculated from metaD~\cite{2011-JCTC-Zhang-Voth}. 
In the WT-metaD simulations, soft walls were added on both side of the interfacial layer to limit sampling inside the AA film.\\

Once representative pathways were identified, we calculated the associated PMF using US, 
with the $Z-$ Cartesian coordinate along the preliminary pathways as collective variable. 
The system was free in the $X-Y$ plane. 
To design the US windows, it is common practice to implement one of two approaches:
 1) the system is dragged using the steered dynamics~\cite{2011-CMS-Kastner} 
 or 2) the initial structures are prepared by changing the position of the CV directly 
 followed by energy minimization~\cite{2013-BJ-Nishizawa-Nishizawa}. 
As discussed by Nishizawa~\cite{2013-BJ-Nishizawa-Nishizawa}, it is possible, 
with these protocols, to drive the system into configurations 
with no reverse transition (\textit{i.e.}, breaking the ergodicity of the system). This can be particularly true 
 with fluid/soft systems, where covalent bonds are not present. This shortcoming can be associated 
to \textit{artifactual} kinetic barriers between consecutive states that slow sampling of configurations during US. 
These effects could render the free-energy profile (FEP) analysis inappropriate and/or unfeasible using a reasonable 
amount of computation.
To avoid these shortcomings, we used the adiabatic biased molecular dynamics (ABMD) 
framework, which is based on the local fluctuations of the system~\cite{1999-JMB-Paci-Karplus,1999-JCP-Marchi-Ballone,
2011-JCP-Camilloni-Tiana,2016-FD-Sicard-Striolo} .

Furthermore, one must avoid inefficient sampling in the CV orthogonal degrees 
of freedom (either $X-$ or $Y-$ direction), despite ensuring good overlap in the $Z-$ direction.
As others indicated~\cite{2004-PNAS-Allen-Roux,2011-JCTC-Zhang-Voth,2012-JCTC-Zhu-Hummer}, 
this numerical issue could cause a discontinuity in configurational space that would introduce PMF errors.
To ensure the results are reliable, we controlled the efficiency of the sampling along the CV orthogonal degrees of freedom 
with the implementation of the \textit{flat-bottomed} potential~\cite{2004-PNAS-Allen-Roux,
2011-JCTC-Zhang-Voth,2012-JCTC-Zhu-Hummer}
\begin{equation}
\label{metaD-flatbottom}
V_{fb}(q) = \frac{1}{2} k_{fb} \Theta\big[\rho(q)-R_{fb}\big]\big(\rho(q)-R_{fb}\big)^2 \, ,
\end{equation}
where $q$ denotes the $X-$ or $Y-$ direction, $\rho(q)$ the distance from the path along which 
the US windows are distributed, $k_{fb}$ the harmonic force constant, $\Theta(q) \equiv 1-\Pi(q)$ 
with $\Pi(q)$ the rectangular function, and $R_{fb}$ the confinement radius. 
No constraint was added in the $X-Y$ plane in the present PMF calculations 
using the $Z$-coordinate as CV, unless the methane molecule is close to the bulk solvent or the hydrate.
Therefore, as $V_{fb}(q)$ is \textit{flat-bottomed}, it has no effect when the methane molecule travels through 
the interfacial layer. This bias only affects the overall offset of the reconstructed FEP in the layer 
and not its shape~\cite{2012-JCTC-Zhu-Hummer}.

Upon completion of the US simulations, FEPs were calculated from the final $4$~ns of simulation time 
using the weighted histogram analysis method (WHAM)~\cite{WHAM}. Statistical error analysis 
was conducted using the integrated Monte Carlo bootstrapping framework~\cite{2010-JCTC-Hub-vanderSpoel}, 
implemented in WHAM~\cite{WHAM}, as discussed in the SI.\\

To estimate the rate of the methane transport across the interfacial layer, 
we extended the \textit{standard} scope of WT-metaD 
considering the recent method or Parrinello, Salvalaglio and Tiwary~\cite{2013-PRL-Tiwary-Parrinello,
2014-JCTC-Salvalaglio-Parrinello}. This technique has been applied to study transition rates 
in biological and chemical systems~\cite{2014-JPCL-Schnelder-Reuter,2015-JCP-Sicard-Manghi,2015-PNAS-Salvalaglio-Parrinello,
2015-PNAS-Tiwary-Parrinello,2015-PNAS-Tiwary-Berne,2016-FD-Piaggi-Parrinello,2017-NC-Bochicchio-Pavan}.
We denote by $\tau$ the \textit{physical} mean transition time for the methane to pass over the energy barrier 
and by $\tau_M$ the mean transition time obtained from the WT-metaD run.
The latter is linked to the \textit{physical} mean 
transition time, $\tau$, by means of the acceleration factor,
\begin{equation}
 \alpha(t) = \tau/\tau_M = \braket{e^{\beta V_{\textrm{bias}}(s,t)}}_M \, ,
\end{equation} 
where the angular brakets 
denote an average over a WT-metaD run confined to a metastable basin, 
and $V_{\textrm{bias}}(s,t)$ is the WT-metaD time-dependent bias defined in Eq.~\ref{metaD-potential}. 
To avoid depositing bias in the transition state region, we increase the time lag between 
two successive Gaussian depositions in the WT-metaD framework, $\tau_G = 400$ ps~\cite{2013-PRL-Tiwary-Parrinello,2014-JCTC-Salvalaglio-Parrinello}.
We  confirmed that the statistics of transition times follows a Poisson distribution, performing a two-sample 
Kolmogorov-Smirnov (KS) test. 
The compliance to the KS test allowed us to asses the reliability of the CV considered to distinguish 
between the different metastable states within the US framework~\cite{2013-PRL-Tiwary-Parrinello,2014-JCTC-Salvalaglio-Parrinello} and to reconstruct accurately the PMFs.

\section*{Acknowledgements}

%We acknowledge T. Bui for his help in setting up the system and for useful discussions. 
The authors thank Matteo Salvalaglio for fruitful discussion concerning the metaD algorithm.
This work was granted to the HPC resources of the ARCHER UK National Supercomputing Service 
(http://www.archer.ac.uk). The authors are grateful for the financial support provided by 
Halliburton and the UK Engineering and Physical Sciences Research Council (EPSRC) under 
grant numbers 527889 and EP/N007123/1.

\bibliography{rsc} %You need to replace "rsc" on this line with the name of your .bib file

\begin{thebibliography}{125}
\expandafter\ifx\csname natexlab\endcsname\relax\def\natexlab#1{#1}\fi
\expandafter\ifx\csname bibnamefont\endcsname\relax
  \def\bibnamefont#1{#1}\fi
\expandafter\ifx\csname bibfnamefont\endcsname\relax
  \def\bibfnamefont#1{#1}\fi
\expandafter\ifx\csname citenamefont\endcsname\relax
  \def\citenamefont#1{#1}\fi
\expandafter\ifx\csname url\endcsname\relax
  \def\url#1{\texttt{#1}}\fi
\expandafter\ifx\csname urlprefix\endcsname\relax\def\urlprefix{URL }\fi
\providecommand{\bibinfo}[2]{#2}
\providecommand{\eprint}[2][]{\url{#2}}

\bibitem[{\citenamefont{Koh et~al.}(2002)\citenamefont{Koh, Westacott, Zhang,
  Hirachand, Creek, and Soper}}]{2002-FPE-Koh-Soper}
\bibinfo{author}{\bibfnamefont{C.}~\bibnamefont{Koh}},
  \bibinfo{author}{\bibfnamefont{R.}~\bibnamefont{Westacott}},
  \bibinfo{author}{\bibfnamefont{W.}~\bibnamefont{Zhang}},
  \bibinfo{author}{\bibfnamefont{K.}~\bibnamefont{Hirachand}},
  \bibinfo{author}{\bibfnamefont{J.}~\bibnamefont{Creek}}, \bibnamefont{and}
  \bibinfo{author}{\bibfnamefont{A.}~\bibnamefont{Soper}},
  \bibinfo{journal}{Fluid Phase Equilib.} \textbf{\bibinfo{volume}{194-197}},
  \bibinfo{pages}{143} (\bibinfo{year}{2002}).

\bibitem[{\citenamefont{Kelland}(2006)}]{2006-EF-Kelland}
\bibinfo{author}{\bibfnamefont{M.}~\bibnamefont{Kelland}},
  \bibinfo{journal}{Energy Fuels} \textbf{\bibinfo{volume}{20}},
  \bibinfo{pages}{825} (\bibinfo{year}{2006}).

\bibitem[{\citenamefont{Sloan and Koh}(2008)}]{2008-CRCPress-Sloan-Koh}
\bibinfo{author}{\bibfnamefont{E.}~\bibnamefont{Sloan}} \bibnamefont{and}
  \bibinfo{author}{\bibfnamefont{C.}~\bibnamefont{Koh}},
  \emph{\bibinfo{title}{Clathrate hydrates of natural gases, 3rd Ed.}}
  (\bibinfo{publisher}{CRC Press: Boca Raton, Florida}, \bibinfo{year}{2008}).

\bibitem[{\citenamefont{Walsh et~al.}(2009)\citenamefont{Walsh, Koh, Sloan,
  Sum, and Wu}}]{2009-Science-Walsh-Wu}
\bibinfo{author}{\bibfnamefont{M.}~\bibnamefont{Walsh}},
  \bibinfo{author}{\bibfnamefont{C.}~\bibnamefont{Koh}},
  \bibinfo{author}{\bibfnamefont{E.}~\bibnamefont{Sloan}},
  \bibinfo{author}{\bibfnamefont{A.}~\bibnamefont{Sum}}, \bibnamefont{and}
  \bibinfo{author}{\bibfnamefont{D.}~\bibnamefont{Wu}},
  \bibinfo{journal}{Science} \textbf{\bibinfo{volume}{326}},
  \bibinfo{pages}{1095} (\bibinfo{year}{2009}).

\bibitem[{\citenamefont{Michalis et~al.}(2015)\citenamefont{Michalis, Costandy,
  Tsimpanogiannis, Stubos, and Economou}}]{2015-JCP-Michalis-Economou}
\bibinfo{author}{\bibfnamefont{V.}~\bibnamefont{Michalis}},
  \bibinfo{author}{\bibfnamefont{J.}~\bibnamefont{Costandy}},
  \bibinfo{author}{\bibfnamefont{I.}~\bibnamefont{Tsimpanogiannis}},
  \bibinfo{author}{\bibfnamefont{A.}~\bibnamefont{Stubos}}, \bibnamefont{and}
  \bibinfo{author}{\bibfnamefont{I.}~\bibnamefont{Economou}},
  \bibinfo{journal}{J. Chem. Phys.} \textbf{\bibinfo{volume}{142}},
  \bibinfo{pages}{044501} (\bibinfo{year}{2015}).

\bibitem[{\citenamefont{Brewer et~al.}(1997)\citenamefont{Brewer, Jr.,
  Friederich, Kvenvolden, Orange, McFarlane, and
  Kirkwood}}]{1997-Geology-Brewer-Kirkwood}
\bibinfo{author}{\bibfnamefont{P.}~\bibnamefont{Brewer}},
  \bibinfo{author}{\bibfnamefont{F.~O.} \bibnamefont{Jr.}},
  \bibinfo{author}{\bibfnamefont{G.}~\bibnamefont{Friederich}},
  \bibinfo{author}{\bibfnamefont{K.}~\bibnamefont{Kvenvolden}},
  \bibinfo{author}{\bibfnamefont{D.}~\bibnamefont{Orange}},
  \bibinfo{author}{\bibfnamefont{J.}~\bibnamefont{McFarlane}},
  \bibnamefont{and} \bibinfo{author}{\bibfnamefont{W.}~\bibnamefont{Kirkwood}},
  \bibinfo{journal}{Geology} \textbf{\bibinfo{volume}{25}},
  \bibinfo{pages}{407} (\bibinfo{year}{1997}).

\bibitem[{\citenamefont{Sum et~al.}(2009)\citenamefont{Sum, Koh, and
  Sloan}}]{2009-IECR-Sum-Sloan}
\bibinfo{author}{\bibfnamefont{A.}~\bibnamefont{Sum}},
  \bibinfo{author}{\bibfnamefont{C.}~\bibnamefont{Koh}}, \bibnamefont{and}
  \bibinfo{author}{\bibfnamefont{E.}~\bibnamefont{Sloan}},
  \bibinfo{journal}{Ind. Eng. Chem. Res.} \textbf{\bibinfo{volume}{48}},
  \bibinfo{pages}{7457} (\bibinfo{year}{2009}).

\bibitem[{\citenamefont{Kaiho et~al.}(1996)\citenamefont{Kaiho, Arinobu,
  Ishiwatari, Morgans, Okada, Takeda, Tazaki, Zhou, Kajiwara, Matsumoto
  et~al.}}]{1996-Paleoceanography-Kaiho-Wada}
\bibinfo{author}{\bibfnamefont{K.}~\bibnamefont{Kaiho}},
  \bibinfo{author}{\bibfnamefont{T.}~\bibnamefont{Arinobu}},
  \bibinfo{author}{\bibfnamefont{R.}~\bibnamefont{Ishiwatari}},
  \bibinfo{author}{\bibfnamefont{H.}~\bibnamefont{Morgans}},
  \bibinfo{author}{\bibfnamefont{H.}~\bibnamefont{Okada}},
  \bibinfo{author}{\bibfnamefont{N.}~\bibnamefont{Takeda}},
  \bibinfo{author}{\bibfnamefont{K.}~\bibnamefont{Tazaki}},
  \bibinfo{author}{\bibfnamefont{G.}~\bibnamefont{Zhou}},
  \bibinfo{author}{\bibfnamefont{Y.}~\bibnamefont{Kajiwara}},
  \bibinfo{author}{\bibfnamefont{R.}~\bibnamefont{Matsumoto}},
  \bibnamefont{et~al.}, \bibinfo{journal}{Paleoceanography}
  \textbf{\bibinfo{volume}{11}}, \bibinfo{pages}{447} (\bibinfo{year}{1996}).

\bibitem[{\citenamefont{Park et~al.}(2006)\citenamefont{Park, Ni, C\^{o}t\'e,
  Choi, Huang, Uribe-Romo, Chae, O’Keeffe, and Yaghi}}]{2006-PNAS-Park-Yaghi}
\bibinfo{author}{\bibfnamefont{K.}~\bibnamefont{Park}},
  \bibinfo{author}{\bibfnamefont{Z.}~\bibnamefont{Ni}},
  \bibinfo{author}{\bibfnamefont{A.}~\bibnamefont{C\^{o}t\'e}},
  \bibinfo{author}{\bibfnamefont{J.}~\bibnamefont{Choi}},
  \bibinfo{author}{\bibfnamefont{R.}~\bibnamefont{Huang}},
  \bibinfo{author}{\bibfnamefont{F.}~\bibnamefont{Uribe-Romo}},
  \bibinfo{author}{\bibfnamefont{H.}~\bibnamefont{Chae}},
  \bibinfo{author}{\bibfnamefont{M.}~\bibnamefont{O’Keeffe}},
  \bibnamefont{and} \bibinfo{author}{\bibfnamefont{O.}~\bibnamefont{Yaghi}},
  \bibinfo{journal}{Proc. Natl. Acad. Sci. U.S.A.}
  \textbf{\bibinfo{volume}{103}}, \bibinfo{pages}{12690}
  (\bibinfo{year}{2006}).

\bibitem[{\citenamefont{Sloan}(2003)}]{2003-Nature-Sloan}
\bibinfo{author}{\bibfnamefont{E.}~\bibnamefont{Sloan}},
  \bibinfo{journal}{Nature} \textbf{\bibinfo{volume}{426}},
  \bibinfo{pages}{353} (\bibinfo{year}{2003}).

\bibitem[{\citenamefont{Florusse et~al.}(2004)\citenamefont{Florusse, Peters,
  Schoonman, Hester, Koh, Dec, Marsh, and Sloan}}]{2004-Science-Florusse-Sloan}
\bibinfo{author}{\bibfnamefont{L.}~\bibnamefont{Florusse}},
  \bibinfo{author}{\bibfnamefont{C.}~\bibnamefont{Peters}},
  \bibinfo{author}{\bibfnamefont{J.}~\bibnamefont{Schoonman}},
  \bibinfo{author}{\bibfnamefont{K.}~\bibnamefont{Hester}},
  \bibinfo{author}{\bibfnamefont{C.}~\bibnamefont{Koh}},
  \bibinfo{author}{\bibfnamefont{S.}~\bibnamefont{Dec}},
  \bibinfo{author}{\bibfnamefont{K.}~\bibnamefont{Marsh}}, \bibnamefont{and}
  \bibinfo{author}{\bibfnamefont{E.}~\bibnamefont{Sloan}},
  \bibinfo{journal}{Science} \textbf{\bibinfo{volume}{306}},
  \bibinfo{pages}{469} (\bibinfo{year}{2004}).

\bibitem[{\citenamefont{Mao et~al.}(2002)\citenamefont{Mao, Mao, Goncharov,
  Struzhkin, Guo, Hu, Shu, Hemley, Somayazulu, and
  Zhao}}]{2002-Science-Mao-Zhao}
\bibinfo{author}{\bibfnamefont{W.}~\bibnamefont{Mao}},
  \bibinfo{author}{\bibfnamefont{H.}~\bibnamefont{Mao}},
  \bibinfo{author}{\bibfnamefont{A.}~\bibnamefont{Goncharov}},
  \bibinfo{author}{\bibfnamefont{V.}~\bibnamefont{Struzhkin}},
  \bibinfo{author}{\bibfnamefont{Q.}~\bibnamefont{Guo}},
  \bibinfo{author}{\bibfnamefont{J.}~\bibnamefont{Hu}},
  \bibinfo{author}{\bibfnamefont{J.}~\bibnamefont{Shu}},
  \bibinfo{author}{\bibfnamefont{R.}~\bibnamefont{Hemley}},
  \bibinfo{author}{\bibfnamefont{M.}~\bibnamefont{Somayazulu}},
  \bibnamefont{and} \bibinfo{author}{\bibfnamefont{Y.}~\bibnamefont{Zhao}},
  \bibinfo{journal}{Science} \textbf{\bibinfo{volume}{297}},
  \bibinfo{pages}{2247} (\bibinfo{year}{2002}).

\bibitem[{\citenamefont{Ogawa et~al.}(2006)\citenamefont{Ogawa, Ito, Watanabe,
  ichi Tahara, Hiraoka, ichi Ochiai, Ohmura, and Mori}}]{2006-ATE-Ogawa-Mori}
\bibinfo{author}{\bibfnamefont{T.}~\bibnamefont{Ogawa}},
  \bibinfo{author}{\bibfnamefont{T.}~\bibnamefont{Ito}},
  \bibinfo{author}{\bibfnamefont{K.}~\bibnamefont{Watanabe}},
  \bibinfo{author}{\bibfnamefont{K.}~\bibnamefont{ichi Tahara}},
  \bibinfo{author}{\bibfnamefont{R.}~\bibnamefont{Hiraoka}},
  \bibinfo{author}{\bibfnamefont{J.}~\bibnamefont{ichi Ochiai}},
  \bibinfo{author}{\bibfnamefont{R.}~\bibnamefont{Ohmura}}, \bibnamefont{and}
  \bibinfo{author}{\bibfnamefont{Y.}~\bibnamefont{Mori}},
  \bibinfo{journal}{Appl. Therm. Eng.} \textbf{\bibinfo{volume}{26}},
  \bibinfo{pages}{2157} (\bibinfo{year}{2006}).

\bibitem[{\citenamefont{Fisher et~al.}(2000)\citenamefont{Fisher, MacDonald,
  Sassen, Young, Macko, Hourdez, Carney, Joye, and
  McMullin}}]{2000-Naturwissenschaften-Fisher-McMullin}
\bibinfo{author}{\bibfnamefont{C.}~\bibnamefont{Fisher}},
  \bibinfo{author}{\bibfnamefont{I.}~\bibnamefont{MacDonald}},
  \bibinfo{author}{\bibfnamefont{R.}~\bibnamefont{Sassen}},
  \bibinfo{author}{\bibfnamefont{C.}~\bibnamefont{Young}},
  \bibinfo{author}{\bibfnamefont{S.}~\bibnamefont{Macko}},
  \bibinfo{author}{\bibfnamefont{S.}~\bibnamefont{Hourdez}},
  \bibinfo{author}{\bibfnamefont{R.}~\bibnamefont{Carney}},
  \bibinfo{author}{\bibfnamefont{S.}~\bibnamefont{Joye}}, \bibnamefont{and}
  \bibinfo{author}{\bibfnamefont{E.}~\bibnamefont{McMullin}},
  \bibinfo{journal}{Naturwissenschaften} \textbf{\bibinfo{volume}{87}},
  \bibinfo{pages}{184} (\bibinfo{year}{2000}).

\bibitem[{\citenamefont{Milton}(1974)}]{1974-Science-Milton}
\bibinfo{author}{\bibfnamefont{D.}~\bibnamefont{Milton}},
  \bibinfo{journal}{Science} \textbf{\bibinfo{volume}{183}},
  \bibinfo{pages}{654} (\bibinfo{year}{1974}).

\bibitem[{\citenamefont{Sloan et~al.}(2011)\citenamefont{Sloan, Koh, Sum,
  Ballard, Creek, Eaton, Lachance, McMullen, Palermo, Shoup
  et~al.}}]{2011-Elsevier-Sloan-Talley}
\bibinfo{author}{\bibfnamefont{E.}~\bibnamefont{Sloan}},
  \bibinfo{author}{\bibfnamefont{C.}~\bibnamefont{Koh}},
  \bibinfo{author}{\bibfnamefont{A.}~\bibnamefont{Sum}},
  \bibinfo{author}{\bibfnamefont{A.}~\bibnamefont{Ballard}},
  \bibinfo{author}{\bibfnamefont{J.}~\bibnamefont{Creek}},
  \bibinfo{author}{\bibfnamefont{M.}~\bibnamefont{Eaton}},
  \bibinfo{author}{\bibfnamefont{J.}~\bibnamefont{Lachance}},
  \bibinfo{author}{\bibfnamefont{N.}~\bibnamefont{McMullen}},
  \bibinfo{author}{\bibfnamefont{T.}~\bibnamefont{Palermo}},
  \bibinfo{author}{\bibfnamefont{G.}~\bibnamefont{Shoup}},
  \bibnamefont{et~al.}, \emph{\bibinfo{title}{Natural Gas Hydrates in Flow
  Assurance, 3rd Ed.}} (\bibinfo{publisher}{Gulf Professional
  Pub./Elsevier:Burlington, Massachusetts}, \bibinfo{year}{2011}).

\bibitem[{\citenamefont{Mullin}(2011)}]{2001-OxfordPress-Mullin}
\bibinfo{author}{\bibfnamefont{J.}~\bibnamefont{Mullin}},
  \emph{\bibinfo{title}{Chapter 5: Nucleation. In Crystallization, 4th Ed.}}
  (\bibinfo{publisher}{Butterworth-Heinemann: Oxford (UK)},
  \bibinfo{year}{2011}).

\bibitem[{\citenamefont{Freer et~al.}(2001)\citenamefont{Freer, Selim, and
  Sloan}}]{2001-FPE-Freer-Sloan}
\bibinfo{author}{\bibfnamefont{E.}~\bibnamefont{Freer}},
  \bibinfo{author}{\bibfnamefont{M.}~\bibnamefont{Selim}}, \bibnamefont{and}
  \bibinfo{author}{\bibfnamefont{E.}~\bibnamefont{Sloan}},
  \bibinfo{journal}{Fluid Phase Equilib.} \textbf{\bibinfo{volume}{185}},
  \bibinfo{pages}{65} (\bibinfo{year}{2001}).

\bibitem[{\citenamefont{Turner}(April 2005)}]{2005-PhD-Turner}
\bibinfo{author}{\bibfnamefont{D.}~\bibnamefont{Turner}},
  \emph{\bibinfo{title}{Clathrate Hydrate Formation in Water-in-oil
  Dispersions}} (\bibinfo{publisher}{Ph.D. Thesis, Colorado School of Mines,
  Golden (CO)}, \bibinfo{year}{April 2005}).

\bibitem[{\citenamefont{Sloan}(2000)}]{2000-RichardsonPress-Sloan}
\bibinfo{author}{\bibfnamefont{E.}~\bibnamefont{Sloan}},
  \emph{\bibinfo{title}{Hydrate Engineering}} (\bibinfo{publisher}{Society of
  Petroleum Engineers, Richardson (TX)}, \bibinfo{year}{2000}).

\bibitem[{\citenamefont{Lee et~al.}(2015)\citenamefont{Lee, Cho, Lee, Linga,
  Kang, and Lee}}]{2015-ET-Lee-Lee}
\bibinfo{author}{\bibfnamefont{J.}~\bibnamefont{Lee}},
  \bibinfo{author}{\bibfnamefont{S.}~\bibnamefont{Cho}},
  \bibinfo{author}{\bibfnamefont{J.}~\bibnamefont{Lee}},
  \bibinfo{author}{\bibfnamefont{P.}~\bibnamefont{Linga}},
  \bibinfo{author}{\bibfnamefont{K.}~\bibnamefont{Kang}}, \bibnamefont{and}
  \bibinfo{author}{\bibfnamefont{J.}~\bibnamefont{Lee}},
  \bibinfo{journal}{Energy Technol.} \textbf{\bibinfo{volume}{3}},
  \bibinfo{pages}{925} (\bibinfo{year}{2015}).

\bibitem[{\citenamefont{Fandino and Ruffine}(2014)}]{2014-Fuel-Fandino-Ruffine}
\bibinfo{author}{\bibfnamefont{O.}~\bibnamefont{Fandino}} \bibnamefont{and}
  \bibinfo{author}{\bibfnamefont{L.}~\bibnamefont{Ruffine}},
  \bibinfo{journal}{Fuel} \textbf{\bibinfo{volume}{117}}, \bibinfo{pages}{442}
  (\bibinfo{year}{2014}).

\bibitem[{\citenamefont{Kulkarni et~al.}(2013)\citenamefont{Kulkarni, Kadam,
  Meekes, Stankiewicz, and ter Horst}}]{2013-CGD-Kulkarni-terHorst}
\bibinfo{author}{\bibfnamefont{S.}~\bibnamefont{Kulkarni}},
  \bibinfo{author}{\bibfnamefont{S.}~\bibnamefont{Kadam}},
  \bibinfo{author}{\bibfnamefont{H.}~\bibnamefont{Meekes}},
  \bibinfo{author}{\bibfnamefont{A.}~\bibnamefont{Stankiewicz}},
  \bibnamefont{and} \bibinfo{author}{\bibfnamefont{J.}~\bibnamefont{ter
  Horst}}, \bibinfo{journal}{Cryst. Growth Des.} \textbf{\bibinfo{volume}{13}},
  \bibinfo{pages}{2435} (\bibinfo{year}{2013}).

\bibitem[{\citenamefont{Abay and Svartaas}(2011)}]{2011-EF-Abay-Svartaas}
\bibinfo{author}{\bibfnamefont{H.}~\bibnamefont{Abay}} \bibnamefont{and}
  \bibinfo{author}{\bibfnamefont{T.}~\bibnamefont{Svartaas}},
  \bibinfo{journal}{Energy Fuels} \textbf{\bibinfo{volume}{25}},
  \bibinfo{pages}{42} (\bibinfo{year}{2011}).

\bibitem[{\citenamefont{Li et~al.}(2014)\citenamefont{Li, Sun, Liu, Li, Chen,
  and Sum}}]{2014-SR-Li-Sum}
\bibinfo{author}{\bibfnamefont{S.}~\bibnamefont{Li}},
  \bibinfo{author}{\bibfnamefont{C.}~\bibnamefont{Sun}},
  \bibinfo{author}{\bibfnamefont{B.}~\bibnamefont{Liu}},
  \bibinfo{author}{\bibfnamefont{Z.}~\bibnamefont{Li}},
  \bibinfo{author}{\bibfnamefont{G.}~\bibnamefont{Chen}}, \bibnamefont{and}
  \bibinfo{author}{\bibfnamefont{A.}~\bibnamefont{Sum}}, \bibinfo{journal}{Sci.
  Rep.} \textbf{\bibinfo{volume}{4}}, \bibinfo{pages}{1}
  (\bibinfo{year}{2014}).

\bibitem[{\citenamefont{Ke and Kelland}(2016)}]{2016-EF-Ke-Kelland}
\bibinfo{author}{\bibfnamefont{W.}~\bibnamefont{Ke}} \bibnamefont{and}
  \bibinfo{author}{\bibfnamefont{M.}~\bibnamefont{Kelland}},
  \bibinfo{journal}{Energy Fuels} \textbf{\bibinfo{volume}{30}},
  \bibinfo{pages}{10015} (\bibinfo{year}{2016}).

\bibitem[{\citenamefont{Yuhara et~al.}(2015)\citenamefont{Yuhara, Barnes, Suh,
  Knott, Beckham, Yasuoka, Wu, and Sum}}]{2015-FD-Yuhara-Sum}
\bibinfo{author}{\bibfnamefont{D.}~\bibnamefont{Yuhara}},
  \bibinfo{author}{\bibfnamefont{B.}~\bibnamefont{Barnes}},
  \bibinfo{author}{\bibfnamefont{D.}~\bibnamefont{Suh}},
  \bibinfo{author}{\bibfnamefont{B.}~\bibnamefont{Knott}},
  \bibinfo{author}{\bibfnamefont{G.}~\bibnamefont{Beckham}},
  \bibinfo{author}{\bibfnamefont{K.}~\bibnamefont{Yasuoka}},
  \bibinfo{author}{\bibfnamefont{D.}~\bibnamefont{Wu}}, \bibnamefont{and}
  \bibinfo{author}{\bibfnamefont{A.}~\bibnamefont{Sum}},
  \bibinfo{journal}{Faraday Discuss.} \textbf{\bibinfo{volume}{179}},
  \bibinfo{pages}{463} (\bibinfo{year}{2015}).

\bibitem[{\citenamefont{Moon et~al.}(2007)\citenamefont{Moon, Hawtin, and
  Rodger}}]{2007-FD-Moon-Rodger}
\bibinfo{author}{\bibfnamefont{C.}~\bibnamefont{Moon}},
  \bibinfo{author}{\bibfnamefont{R.}~\bibnamefont{Hawtin}}, \bibnamefont{and}
  \bibinfo{author}{\bibfnamefont{P.}~\bibnamefont{Rodger}},
  \bibinfo{journal}{Faraday Discuss.} \textbf{\bibinfo{volume}{136}},
  \bibinfo{pages}{367} (\bibinfo{year}{2007}).

\bibitem[{\citenamefont{Vatamanu and
  Kusalik}(2006)}]{2006-JPCB-Vatamanu-Kusalik}
\bibinfo{author}{\bibfnamefont{J.}~\bibnamefont{Vatamanu}} \bibnamefont{and}
  \bibinfo{author}{\bibfnamefont{P.}~\bibnamefont{Kusalik}},
  \bibinfo{journal}{J. Phys. Chem. B} \textbf{\bibinfo{volume}{26}},
  \bibinfo{pages}{15896} (\bibinfo{year}{2006}).

\bibitem[{\citenamefont{Sum et~al.}(2012)\citenamefont{Sum, Koh, and
  Sloan}}]{2012-EF-Sum-Sloan}
\bibinfo{author}{\bibfnamefont{A.}~\bibnamefont{Sum}},
  \bibinfo{author}{\bibfnamefont{C.}~\bibnamefont{Koh}}, \bibnamefont{and}
  \bibinfo{author}{\bibfnamefont{E.}~\bibnamefont{Sloan}},
  \bibinfo{journal}{Energy Fuels} \textbf{\bibinfo{volume}{26}},
  \bibinfo{pages}{4046} (\bibinfo{year}{2012}).

\bibitem[{\citenamefont{Jr. and Large}(2008)}]{2008-CES-Ribeiro-Large}
\bibinfo{author}{\bibfnamefont{C.~R.} \bibnamefont{Jr.}} \bibnamefont{and}
  \bibinfo{author}{\bibfnamefont{P.}~\bibnamefont{Large}},
  \bibinfo{journal}{Chem. Eng. Sci.} \textbf{\bibinfo{volume}{63}},
  \bibinfo{pages}{2007} (\bibinfo{year}{2008}).

\bibitem[{\citenamefont{Zhang et~al.}(2014)\citenamefont{Zhang, Zhang, Huang,
  QI, Zhang, Ren, Wu, and Fang}}]{2014-PED-Zhang-Fang}
\bibinfo{author}{\bibfnamefont{L.}~\bibnamefont{Zhang}},
  \bibinfo{author}{\bibfnamefont{C.}~\bibnamefont{Zhang}},
  \bibinfo{author}{\bibfnamefont{H.}~\bibnamefont{Huang}},
  \bibinfo{author}{\bibfnamefont{D.}~\bibnamefont{QI}},
  \bibinfo{author}{\bibfnamefont{Y.}~\bibnamefont{Zhang}},
  \bibinfo{author}{\bibfnamefont{S.}~\bibnamefont{Ren}},
  \bibinfo{author}{\bibfnamefont{Z.}~\bibnamefont{Wu}}, \bibnamefont{and}
  \bibinfo{author}{\bibfnamefont{M.}~\bibnamefont{Fang}},
  \bibinfo{journal}{Petrol. Explor. Develop.} \textbf{\bibinfo{volume}{41}},
  \bibinfo{pages}{824} (\bibinfo{year}{2014}).

\bibitem[{\citenamefont{Lederhos et~al.}(1996)\citenamefont{Lederhos, Long,
  Sum, Christiansen, and Sloan}}]{1996-CES-Lederhos-Sloan}
\bibinfo{author}{\bibfnamefont{J.}~\bibnamefont{Lederhos}},
  \bibinfo{author}{\bibfnamefont{J.}~\bibnamefont{Long}},
  \bibinfo{author}{\bibfnamefont{A.}~\bibnamefont{Sum}},
  \bibinfo{author}{\bibfnamefont{R.}~\bibnamefont{Christiansen}},
  \bibnamefont{and} \bibinfo{author}{\bibfnamefont{E.}~\bibnamefont{Sloan}},
  \bibinfo{journal}{Chem. Eng. Sci.} \textbf{\bibinfo{volume}{51}},
  \bibinfo{pages}{1221} (\bibinfo{year}{1996}).

\bibitem[{\citenamefont{Zhao et~al.}(2016)\citenamefont{Zhao, Sun, and
  A.Firoozabadi}}]{2016-Fuel-Zhao-Firoozabadi}
\bibinfo{author}{\bibfnamefont{H.}~\bibnamefont{Zhao}},
  \bibinfo{author}{\bibfnamefont{M.}~\bibnamefont{Sun}}, \bibnamefont{and}
  \bibinfo{author}{\bibnamefont{A.Firoozabadi}}, \bibinfo{journal}{Fuel}
  \textbf{\bibinfo{volume}{180}}, \bibinfo{pages}{187} (\bibinfo{year}{2016}).

\bibitem[{\citenamefont{Sloan}(May 25, 1994)}]{1995-Patent-Sloan}
\bibinfo{author}{\bibfnamefont{E.}~\bibnamefont{Sloan}}, \bibinfo{journal}{U.S.
  Patent US5880319 A}  (\bibinfo{year}{May 25, 1994}).

\bibitem[{\citenamefont{Bui et~al.}(2017)\citenamefont{Bui, Phan, Monteiro,
  Lan, Ceglio, Acosta, Krishnamurthy, and Striolo}}]{2017-Langmuir-Bui-Striolo}
\bibinfo{author}{\bibfnamefont{T.}~\bibnamefont{Bui}},
  \bibinfo{author}{\bibfnamefont{A.}~\bibnamefont{Phan}},
  \bibinfo{author}{\bibfnamefont{D.}~\bibnamefont{Monteiro}},
  \bibinfo{author}{\bibfnamefont{Q.}~\bibnamefont{Lan}},
  \bibinfo{author}{\bibfnamefont{M.}~\bibnamefont{Ceglio}},
  \bibinfo{author}{\bibfnamefont{E.}~\bibnamefont{Acosta}},
  \bibinfo{author}{\bibfnamefont{P.}~\bibnamefont{Krishnamurthy}},
  \bibnamefont{and} \bibinfo{author}{\bibfnamefont{A.}~\bibnamefont{Striolo}},
  \bibinfo{journal}{Langmuir} \textbf{\bibinfo{volume}{33}},
  \bibinfo{pages}{2263} (\bibinfo{year}{2017}).

\bibitem[{\citenamefont{Kelland et~al.}(2009)\citenamefont{Kelland, Svartaas,
  and Andersen}}]{2009-JPSE-Kelland-Andersen}
\bibinfo{author}{\bibfnamefont{M.}~\bibnamefont{Kelland}},
  \bibinfo{author}{\bibfnamefont{T.}~\bibnamefont{Svartaas}}, \bibnamefont{and}
  \bibinfo{author}{\bibfnamefont{L.}~\bibnamefont{Andersen}},
  \bibinfo{journal}{J. Pet. Sci. Eng.} \textbf{\bibinfo{volume}{64}},
  \bibinfo{pages}{1} (\bibinfo{year}{2009}).

\bibitem[{\citenamefont{Aman et~al.}(2014)\citenamefont{Aman, Sloan, Sum, and
  Koh}}]{2014-PCCP-Aman-Koh}
\bibinfo{author}{\bibfnamefont{Z.}~\bibnamefont{Aman}},
  \bibinfo{author}{\bibfnamefont{E.}~\bibnamefont{Sloan}},
  \bibinfo{author}{\bibfnamefont{A.}~\bibnamefont{Sum}}, \bibnamefont{and}
  \bibinfo{author}{\bibfnamefont{C.}~\bibnamefont{Koh}},
  \bibinfo{journal}{Phys. Chem. Chem. Phys.} \textbf{\bibinfo{volume}{16}},
  \bibinfo{pages}{25121} (\bibinfo{year}{2014}).

\bibitem[{\citenamefont{mechanisms responsible for hydrate anti-agglomerant
  performance}(2016)}]{2016-PCCP-Phan-Striolo}
\bibinfo{author}{\bibfnamefont{M.}~\bibnamefont{mechanisms responsible for
  hydrate anti-agglomerant performance}}, \bibinfo{journal}{Phys. Chem. Chem.
  Phys.} \textbf{\bibinfo{volume}{18}}, \bibinfo{pages}{24859}
  (\bibinfo{year}{2016}).

\bibitem[{\citenamefont{Bui et~al.}(2018)\citenamefont{Bui, Sicard, Monteiro,
  Lan, Ceglio, Burress, and Striolo}}]{2017-preprint-Bui-Striolo}
\bibinfo{author}{\bibfnamefont{T.}~\bibnamefont{Bui}},
  \bibinfo{author}{\bibfnamefont{F.}~\bibnamefont{Sicard}},
  \bibinfo{author}{\bibfnamefont{D.}~\bibnamefont{Monteiro}},
  \bibinfo{author}{\bibfnamefont{Q.}~\bibnamefont{Lan}},
  \bibinfo{author}{\bibfnamefont{M.}~\bibnamefont{Ceglio}},
  \bibinfo{author}{\bibfnamefont{C.}~\bibnamefont{Burress}}, \bibnamefont{and}
  \bibinfo{author}{\bibfnamefont{A.}~\bibnamefont{Striolo}},
  \bibinfo{journal}{Nano Lett., submitted for publication}
  (\bibinfo{year}{2018}).

\bibitem[{\citenamefont{Zanota et~al.}(2005)\citenamefont{Zanota, Dicharry, and
  Graciaa}}]{2005-EF-Zanota-Graciaa}
\bibinfo{author}{\bibfnamefont{M.}~\bibnamefont{Zanota}},
  \bibinfo{author}{\bibfnamefont{C.}~\bibnamefont{Dicharry}}, \bibnamefont{and}
  \bibinfo{author}{\bibfnamefont{A.}~\bibnamefont{Graciaa}},
  \bibinfo{journal}{Energy Fuels} \textbf{\bibinfo{volume}{19}},
  \bibinfo{pages}{584} (\bibinfo{year}{2005}).

\bibitem[{\citenamefont{Chen et~al.}(2013)\citenamefont{Chen, Sun, Peng, Liu,
  Si, and et~al.}}]{2013-EF-Chen-etal}
\bibinfo{author}{\bibfnamefont{J.}~\bibnamefont{Chen}},
  \bibinfo{author}{\bibfnamefont{C.}~\bibnamefont{Sun}},
  \bibinfo{author}{\bibfnamefont{B.}~\bibnamefont{Peng}},
  \bibinfo{author}{\bibfnamefont{B.}~\bibnamefont{Liu}},
  \bibinfo{author}{\bibfnamefont{S.}~\bibnamefont{Si}}, \bibnamefont{and}
  \bibinfo{author}{\bibfnamefont{M.~J.} \bibnamefont{et~al.}},
  \bibinfo{journal}{Energy Fuels} \textbf{\bibinfo{volume}{27}},
  \bibinfo{pages}{2488} (\bibinfo{year}{2013}).

\bibitem[{\citenamefont{Kelland et~al.}(2006)\citenamefont{Kelland, Svartaas,
  Ovsthus, Tomita, and Chosa}}]{2006-CES-Kelland-Chosa}
\bibinfo{author}{\bibfnamefont{M.}~\bibnamefont{Kelland}},
  \bibinfo{author}{\bibfnamefont{T.}~\bibnamefont{Svartaas}},
  \bibinfo{author}{\bibfnamefont{J.}~\bibnamefont{Ovsthus}},
  \bibinfo{author}{\bibfnamefont{T.}~\bibnamefont{Tomita}}, \bibnamefont{and}
  \bibinfo{author}{\bibfnamefont{J.}~\bibnamefont{Chosa}},
  \bibinfo{journal}{Chem. Eng. Sci.} \textbf{\bibinfo{volume}{61}},
  \bibinfo{pages}{4048} (\bibinfo{year}{2006}).

\bibitem[{\citenamefont{Sun et~al.}(2012)\citenamefont{Sun, Wang, and
  Firoozabadi}}]{2012-EF-Sun-Firoozabadi}
\bibinfo{author}{\bibfnamefont{M.}~\bibnamefont{Sun}},
  \bibinfo{author}{\bibfnamefont{Y.}~\bibnamefont{Wang}}, \bibnamefont{and}
  \bibinfo{author}{\bibfnamefont{A.}~\bibnamefont{Firoozabadi}},
  \bibinfo{journal}{Energy Fuels} \textbf{\bibinfo{volume}{26}},
  \bibinfo{pages}{5626} (\bibinfo{year}{2012}).

\bibitem[{\citenamefont{Chen et~al.}(2014)\citenamefont{Chen, Liu, Sun, Jia,
  and et~al.}}]{2014-ECM-Chen-etal}
\bibinfo{author}{\bibfnamefont{J.}~\bibnamefont{Chen}},
  \bibinfo{author}{\bibfnamefont{J.}~\bibnamefont{Liu}},
  \bibinfo{author}{\bibfnamefont{C.}~\bibnamefont{Sun}},
  \bibinfo{author}{\bibfnamefont{M.}~\bibnamefont{Jia}}, \bibnamefont{and}
  \bibinfo{author}{\bibfnamefont{B.~L.} \bibnamefont{et~al.}},
  \bibinfo{journal}{Energy Convers. Manage.} \textbf{\bibinfo{volume}{86}},
  \bibinfo{pages}{886} (\bibinfo{year}{2014}).

\bibitem[{\citenamefont{Chen et~al.}(2015)\citenamefont{Chen, Wang, Sun, Li,
  Ren, Jia, Yan, Lv, Liu, and Chen}}]{2015-EF-Chen-etal}
\bibinfo{author}{\bibfnamefont{J.}~\bibnamefont{Chen}},
  \bibinfo{author}{\bibfnamefont{Y.}~\bibnamefont{Wang}},
  \bibinfo{author}{\bibfnamefont{C.}~\bibnamefont{Sun}},
  \bibinfo{author}{\bibfnamefont{F.}~\bibnamefont{Li}},
  \bibinfo{author}{\bibfnamefont{N.}~\bibnamefont{Ren}},
  \bibinfo{author}{\bibfnamefont{M.}~\bibnamefont{Jia}},
  \bibinfo{author}{\bibfnamefont{K.}~\bibnamefont{Yan}},
  \bibinfo{author}{\bibfnamefont{Y.-N.} \bibnamefont{Lv}},
  \bibinfo{author}{\bibfnamefont{B.}~\bibnamefont{Liu}}, \bibnamefont{and}
  \bibinfo{author}{\bibfnamefont{G.-J.} \bibnamefont{Chen}},
  \bibinfo{journal}{Energy Fuels} \textbf{\bibinfo{volume}{29}},
  \bibinfo{pages}{122} (\bibinfo{year}{2015}).

\bibitem[{\citenamefont{Gao}(2009)}]{2009-EF-Gao}
\bibinfo{author}{\bibfnamefont{S.}~\bibnamefont{Gao}}, \bibinfo{journal}{Energy
  Fuels} \textbf{\bibinfo{volume}{23}}, \bibinfo{pages}{2118}
  (\bibinfo{year}{2009}).

\bibitem[{\citenamefont{Kelland}(2014)}]{2014-CRCPress-Kelland}
\bibinfo{author}{\bibfnamefont{M.}~\bibnamefont{Kelland}},
  \emph{\bibinfo{title}{Production Chemicals for the Oil and Gas Industry, 2nd
  Ed.}} (\bibinfo{publisher}{CRC Press: Boca Raton, Florida},
  \bibinfo{year}{2014}).

\bibitem[{\citenamefont{Lauricella et~al.}(2014)\citenamefont{Lauricella,
  Meloni, English, Peters, and Ciccotti}}]{2014-JPCC-Lauricella-Ciccotti}
\bibinfo{author}{\bibfnamefont{M.}~\bibnamefont{Lauricella}},
  \bibinfo{author}{\bibfnamefont{S.}~\bibnamefont{Meloni}},
  \bibinfo{author}{\bibfnamefont{N.}~\bibnamefont{English}},
  \bibinfo{author}{\bibfnamefont{B.}~\bibnamefont{Peters}}, \bibnamefont{and}
  \bibinfo{author}{\bibfnamefont{G.}~\bibnamefont{Ciccotti}},
  \bibinfo{journal}{J. Phys. Chem. C} \textbf{\bibinfo{volume}{118}},
  \bibinfo{pages}{22847} (\bibinfo{year}{2014}).

\bibitem[{\citenamefont{Lauricella et~al.}(2015)\citenamefont{Lauricella,
  Meloni, Liang, English, Kusalik, and
  Ciccotti}}]{2015-JCP-Lauricella-Ciccotti}
\bibinfo{author}{\bibfnamefont{M.}~\bibnamefont{Lauricella}},
  \bibinfo{author}{\bibfnamefont{S.}~\bibnamefont{Meloni}},
  \bibinfo{author}{\bibfnamefont{S.}~\bibnamefont{Liang}},
  \bibinfo{author}{\bibfnamefont{N.}~\bibnamefont{English}},
  \bibinfo{author}{\bibfnamefont{P.}~\bibnamefont{Kusalik}}, \bibnamefont{and}
  \bibinfo{author}{\bibfnamefont{G.}~\bibnamefont{Ciccotti}},
  \bibinfo{journal}{J. Chem. Phys.} \textbf{\bibinfo{volume}{142}},
  \bibinfo{pages}{244503} (\bibinfo{year}{2015}).

\bibitem[{\citenamefont{Hall et~al.}(2016)\citenamefont{Hall, Carpendale, and
  Kusalik}}]{2016-PNAS-Hall-Kusalik}
\bibinfo{author}{\bibfnamefont{K.}~\bibnamefont{Hall}},
  \bibinfo{author}{\bibfnamefont{S.}~\bibnamefont{Carpendale}},
  \bibnamefont{and} \bibinfo{author}{\bibfnamefont{P.}~\bibnamefont{Kusalik}},
  \bibinfo{journal}{Proc. Nat. Acad. Soc. U.S.A.}
  \textbf{\bibinfo{volume}{113}}, \bibinfo{pages}{12041}
  (\bibinfo{year}{2016}).

\bibitem[{\citenamefont{Jimenez-Angeles and Firoozabadi}(June 25-30,
  2017)}]{ICGH9-1}
\bibinfo{author}{\bibfnamefont{F.}~\bibnamefont{Jimenez-Angeles}}
  \bibnamefont{and}
  \bibinfo{author}{\bibfnamefont{A.}~\bibnamefont{Firoozabadi}},
  \bibinfo{journal}{Proceeding of the 9th International Conference on Gas
  Hydrates, Denver, Colorado, USA}  (\bibinfo{year}{June 25-30, 2017}).

\bibitem[{\citenamefont{Bellucci et~al.}(June 25-30,
  2017)\citenamefont{Bellucci, Walsh, and Trout}}]{ICGH9-2}
\bibinfo{author}{\bibfnamefont{M.}~\bibnamefont{Bellucci}},
  \bibinfo{author}{\bibfnamefont{M.}~\bibnamefont{Walsh}}, \bibnamefont{and}
  \bibinfo{author}{\bibfnamefont{B.}~\bibnamefont{Trout}},
  \bibinfo{journal}{Proceeding of the 9th International Conference on Gas
  Hydrates, Denver, Colorado, USA}  (\bibinfo{year}{June 25-30, 2017}).

\bibitem[{\citenamefont{Laio and Parrinello}(2002)}]{2002-PNAS-Laio-Parrinello}
\bibinfo{author}{\bibfnamefont{A.}~\bibnamefont{Laio}} \bibnamefont{and}
  \bibinfo{author}{\bibfnamefont{M.}~\bibnamefont{Parrinello}},
  \bibinfo{journal}{Proc. Nat. Acad. Soc. U.S.A.}
  \textbf{\bibinfo{volume}{99}}, \bibinfo{pages}{12562–12566}
  (\bibinfo{year}{2002}).

\bibitem[{\citenamefont{Torrie and
  J.P.Valleau}(1977)}]{1977-JCP-Torrie-Valleau}
\bibinfo{author}{\bibfnamefont{G.}~\bibnamefont{Torrie}} \bibnamefont{and}
  \bibinfo{author}{\bibnamefont{J.P.Valleau}}, \bibinfo{journal}{J. Comput.
  Phys.} \textbf{\bibinfo{volume}{23}}, \bibinfo{pages}{187}
  (\bibinfo{year}{1977}).

\bibitem[{\citenamefont{Abraham et~al.}(2015)\citenamefont{Abraham, Murtola,
  Schulz, P\'all, Smith, Hess, and Lindahl}}]{2015-SoftwareX-Abraham-Lindahl}
\bibinfo{author}{\bibfnamefont{M.}~\bibnamefont{Abraham}},
  \bibinfo{author}{\bibfnamefont{T.}~\bibnamefont{Murtola}},
  \bibinfo{author}{\bibfnamefont{R.}~\bibnamefont{Schulz}},
  \bibinfo{author}{\bibfnamefont{S.}~\bibnamefont{P\'all}},
  \bibinfo{author}{\bibfnamefont{J.}~\bibnamefont{Smith}},
  \bibinfo{author}{\bibfnamefont{B.}~\bibnamefont{Hess}}, \bibnamefont{and}
  \bibinfo{author}{\bibfnamefont{E.}~\bibnamefont{Lindahl}},
  \bibinfo{journal}{SoftwareX} \textbf{\bibinfo{volume}{1-2}},
  \bibinfo{pages}{19} (\bibinfo{year}{2015}).

\bibitem[{\citenamefont{Limongelli et~al.}(2012)\citenamefont{Limongelli,
  Marinelli, Cosconati, Motta, Sartini, Mugnaini, Settimo, Novellino, and
  Parrinello}}]{2012-PNAS-Limongelli-Parrinello}
\bibinfo{author}{\bibfnamefont{V.}~\bibnamefont{Limongelli}},
  \bibinfo{author}{\bibfnamefont{L.}~\bibnamefont{Marinelli}},
  \bibinfo{author}{\bibfnamefont{S.}~\bibnamefont{Cosconati}},
  \bibinfo{author}{\bibfnamefont{C.~L.} \bibnamefont{Motta}},
  \bibinfo{author}{\bibfnamefont{S.}~\bibnamefont{Sartini}},
  \bibinfo{author}{\bibfnamefont{L.}~\bibnamefont{Mugnaini}},
  \bibinfo{author}{\bibfnamefont{F.~D.} \bibnamefont{Settimo}},
  \bibinfo{author}{\bibfnamefont{E.}~\bibnamefont{Novellino}},
  \bibnamefont{and}
  \bibinfo{author}{\bibfnamefont{M.}~\bibnamefont{Parrinello}},
  \bibinfo{journal}{Proc. Natl. Acad. Sci. U.S.A.}
  \textbf{\bibinfo{volume}{109}}, \bibinfo{pages}{1467} (\bibinfo{year}{2012}).

\bibitem[{\citenamefont{Tiwary et~al.}(2015{\natexlab{a}})\citenamefont{Tiwary,
  Mondal, Morrone, and Berne}}]{2015-PNAS-Tiwary-Berne}
\bibinfo{author}{\bibfnamefont{P.}~\bibnamefont{Tiwary}},
  \bibinfo{author}{\bibfnamefont{J.}~\bibnamefont{Mondal}},
  \bibinfo{author}{\bibfnamefont{J.}~\bibnamefont{Morrone}}, \bibnamefont{and}
  \bibinfo{author}{\bibfnamefont{B.}~\bibnamefont{Berne}},
  \bibinfo{journal}{Proc. Nat. Acad. Soc. U.S.A.}
  \textbf{\bibinfo{volume}{112}}, \bibinfo{pages}{12015}
  (\bibinfo{year}{2015}{\natexlab{a}}).

\bibitem[{\citenamefont{Yang et~al.}(2016)\citenamefont{Yang, Bonomi, Calero,
  and Martí}}]{2016-PCCP-Yang-Marti}
\bibinfo{author}{\bibfnamefont{J.}~\bibnamefont{Yang}},
  \bibinfo{author}{\bibfnamefont{M.}~\bibnamefont{Bonomi}},
  \bibinfo{author}{\bibfnamefont{C.}~\bibnamefont{Calero}}, \bibnamefont{and}
  \bibinfo{author}{\bibfnamefont{J.}~\bibnamefont{Martí}},
  \bibinfo{journal}{Phys. Chem. Chem. Phys.} \textbf{\bibinfo{volume}{18}},
  \bibinfo{pages}{9036} (\bibinfo{year}{2016}).

\bibitem[{\citenamefont{Bhakat and
  Soderhjelm}(2017)}]{2017-JCAMD-Bhakat-Soderhjelm}
\bibinfo{author}{\bibfnamefont{S.}~\bibnamefont{Bhakat}} \bibnamefont{and}
  \bibinfo{author}{\bibfnamefont{P.}~\bibnamefont{Soderhjelm}},
  \bibinfo{journal}{J. Comput. Aided Mol. Des.} \textbf{\bibinfo{volume}{31}},
  \bibinfo{pages}{119} (\bibinfo{year}{2017}).

\bibitem[{\citenamefont{Granick}(1992)}]{1992-SPRINGER-Granick}
\bibinfo{author}{\bibfnamefont{S.}~\bibnamefont{Granick}},
  \emph{\bibinfo{title}{Fundamentals of Friction, Macroscopic and Microscopic
  Processes, 3rd Ed.}} (\bibinfo{publisher}{I.L. Singer and H. Pollock Eds.;
  Springer Netherlands}, \bibinfo{year}{1992}).

\bibitem[{\citenamefont{Kramers}(1940)}]{1940-Physica-Kramers}
\bibinfo{author}{\bibfnamefont{H.}~\bibnamefont{Kramers}},
  \bibinfo{journal}{Physica (Amsterdam)} \textbf{\bibinfo{volume}{7}},
  \bibinfo{pages}{284} (\bibinfo{year}{1940}).

\bibitem[{\citenamefont{Hanggi et~al.}(1990)\citenamefont{Hanggi, Talkner, and
  Borkovec}}]{1990-RMP-Hanggi-Borkovec}
\bibinfo{author}{\bibfnamefont{P.}~\bibnamefont{Hanggi}},
  \bibinfo{author}{\bibfnamefont{P.}~\bibnamefont{Talkner}}, \bibnamefont{and}
  \bibinfo{author}{\bibfnamefont{M.}~\bibnamefont{Borkovec}},
  \bibinfo{journal}{Rev. Mod. Phys.} \textbf{\bibinfo{volume}{82}},
  \bibinfo{pages}{251} (\bibinfo{year}{1990}).

\bibitem[{\citenamefont{Pollak and Talkner}(2005)}]{2005-Chaos-Pollak-Talkner}
\bibinfo{author}{\bibfnamefont{E.}~\bibnamefont{Pollak}} \bibnamefont{and}
  \bibinfo{author}{\bibfnamefont{P.}~\bibnamefont{Talkner}},
  \bibinfo{journal}{Chaos} \textbf{\bibinfo{volume}{15}},
  \bibinfo{pages}{026116} (\bibinfo{year}{2005}).

\bibitem[{\citenamefont{Zhou}(2010)}]{2010-QRB-Zhou}
\bibinfo{author}{\bibfnamefont{H.-X.} \bibnamefont{Zhou}},
  \bibinfo{journal}{Quaterly Rev.Biophys.} \textbf{\bibinfo{volume}{43}},
  \bibinfo{pages}{219} (\bibinfo{year}{2010}).

\bibitem[{\citenamefont{Peters}(2017)}]{2017-ELSEVIER-Peters}
\bibinfo{author}{\bibfnamefont{B.}~\bibnamefont{Peters}},
  \emph{\bibinfo{title}{Reaction Rate Theory and Rare Events, 1st Ed.}}
  (\bibinfo{publisher}{Elsevier: Amsterdam, The Netherlands},
  \bibinfo{year}{2017}).

\bibitem[{\citenamefont{Sicard}(2018)}]{2018-arXiv-Sicard}
\bibinfo{author}{\bibfnamefont{F.}~\bibnamefont{Sicard}},
  \bibinfo{journal}{arXiv:1803.03490 [cond-mat.stat-mech]}
  (\bibinfo{year}{2018}).

\bibitem[{\citenamefont{Coffey and Kalmykov}(2012)}]{2012-WS-Coffey-Kalmykov}
\bibinfo{author}{\bibfnamefont{W.}~\bibnamefont{Coffey}} \bibnamefont{and}
  \bibinfo{author}{\bibfnamefont{Y.}~\bibnamefont{Kalmykov}},
  \emph{\bibinfo{title}{The Langevin Equation: With Applications to Stochastic
  Problems in Physics, Chemistry and Electrical Engineering, 3rd Ed.; World
  Scientific Series in Contemporary Chemical Physics}},
  vol.~\bibinfo{volume}{27} (\bibinfo{publisher}{World Scientific Publishing
  Company: Singapore}, \bibinfo{year}{2012}).

\bibitem[{\citenamefont{Marini et~al.}(2008)\citenamefont{Marini, Marconi,
  Rondoni, and Vulpiani}}]{2008-PR-Marini-Vulpiani}
\bibinfo{author}{\bibfnamefont{U.}~\bibnamefont{Marini}},
  \bibinfo{author}{\bibfnamefont{B.}~\bibnamefont{Marconi}},
  \bibinfo{author}{\bibfnamefont{A.~P.~L.} \bibnamefont{Rondoni}},
  \bibnamefont{and} \bibinfo{author}{\bibfnamefont{A.}~\bibnamefont{Vulpiani}},
  \bibinfo{journal}{Phys. Rep.} \textbf{\bibinfo{volume}{461}},
  \bibinfo{pages}{111} (\bibinfo{year}{2008}).

\bibitem[{\citenamefont{Yu et~al.}(2012)\citenamefont{Yu, Gupta, Liu, Neupane,
  Brigley, Sosova, and Woodside}}]{2012-PNAS-Yu-Woodside}
\bibinfo{author}{\bibfnamefont{H.}~\bibnamefont{Yu}},
  \bibinfo{author}{\bibfnamefont{A.}~\bibnamefont{Gupta}},
  \bibinfo{author}{\bibfnamefont{X.}~\bibnamefont{Liu}},
  \bibinfo{author}{\bibfnamefont{K.}~\bibnamefont{Neupane}},
  \bibinfo{author}{\bibfnamefont{A.}~\bibnamefont{Brigley}},
  \bibinfo{author}{\bibfnamefont{I.}~\bibnamefont{Sosova}}, \bibnamefont{and}
  \bibinfo{author}{\bibfnamefont{M.}~\bibnamefont{Woodside}},
  \bibinfo{journal}{Proc. Nat. Acad. Sci. U.S.A.}
  \textbf{\bibinfo{volume}{109}}, \bibinfo{pages}{14452}
  (\bibinfo{year}{2012}).

\bibitem[{\citenamefont{Sicard et~al.}(2015)\citenamefont{Sicard, Destainville,
  and Manghi}}]{2015-JCP-Sicard-Manghi}
\bibinfo{author}{\bibfnamefont{F.}~\bibnamefont{Sicard}},
  \bibinfo{author}{\bibfnamefont{N.}~\bibnamefont{Destainville}},
  \bibnamefont{and} \bibinfo{author}{\bibfnamefont{M.}~\bibnamefont{Manghi}},
  \bibinfo{journal}{J. Chem. Phys.} \textbf{\bibinfo{volume}{142}},
  \bibinfo{pages}{034903} (\bibinfo{year}{2015}).

\bibitem[{\citenamefont{Tiwary and
  Parrinello}(2013)}]{2013-PRL-Tiwary-Parrinello}
\bibinfo{author}{\bibfnamefont{P.}~\bibnamefont{Tiwary}} \bibnamefont{and}
  \bibinfo{author}{\bibfnamefont{M.}~\bibnamefont{Parrinello}},
  \bibinfo{journal}{Phys. Rev. Lett.} \textbf{\bibinfo{volume}{111}},
  \bibinfo{pages}{230602} (\bibinfo{year}{2013}).

\bibitem[{\citenamefont{Salvalaglio et~al.}(2014)\citenamefont{Salvalaglio,
  Tiwary, and Parrinello}}]{2014-JCTC-Salvalaglio-Parrinello}
\bibinfo{author}{\bibfnamefont{M.}~\bibnamefont{Salvalaglio}},
  \bibinfo{author}{\bibfnamefont{P.}~\bibnamefont{Tiwary}}, \bibnamefont{and}
  \bibinfo{author}{\bibfnamefont{M.}~\bibnamefont{Parrinello}},
  \bibinfo{journal}{J. Chem. Theory Comput.} \textbf{\bibinfo{volume}{10}},
  \bibinfo{pages}{1420} (\bibinfo{year}{2014}).

\bibitem[{\citenamefont{Sun et~al.}(2010)\citenamefont{Sun, Peng, Dandekar, Ma,
  and Chen}}]{2010-ARPCSC-Sun-Chen}
\bibinfo{author}{\bibfnamefont{C.}~\bibnamefont{Sun}},
  \bibinfo{author}{\bibfnamefont{B.}~\bibnamefont{Peng}},
  \bibinfo{author}{\bibfnamefont{A.}~\bibnamefont{Dandekar}},
  \bibinfo{author}{\bibfnamefont{Q.}~\bibnamefont{Ma}}, \bibnamefont{and}
  \bibinfo{author}{\bibfnamefont{G.}~\bibnamefont{Chen}},
  \bibinfo{journal}{Annu. Rep. Prog. Chem., Sect. C}
  \textbf{\bibinfo{volume}{106}}, \bibinfo{pages}{77} (\bibinfo{year}{2010}).

\bibitem[{\citenamefont{Abascal et~al.}(2005)\citenamefont{Abascal, Sanz,
  Garcia, and Vega}}]{2005-JCP-Abascal-Vega}
\bibinfo{author}{\bibfnamefont{J.}~\bibnamefont{Abascal}},
  \bibinfo{author}{\bibfnamefont{E.}~\bibnamefont{Sanz}},
  \bibinfo{author}{\bibfnamefont{R.~F.} \bibnamefont{Garcia}},
  \bibnamefont{and} \bibinfo{author}{\bibfnamefont{C.}~\bibnamefont{Vega}},
  \bibinfo{journal}{J. Chem. Phys.} \textbf{\bibinfo{volume}{122}},
  \bibinfo{pages}{234511} (\bibinfo{year}{2005}).

\bibitem[{\citenamefont{Tribello et~al.}(2014)\citenamefont{Tribello, Bonomi,
  Branduardi, Camilloni, and Bussi}}]{2014-CPC-Tribello-Bussi}
\bibinfo{author}{\bibfnamefont{G.}~\bibnamefont{Tribello}},
  \bibinfo{author}{\bibfnamefont{M.}~\bibnamefont{Bonomi}},
  \bibinfo{author}{\bibfnamefont{D.}~\bibnamefont{Branduardi}},
  \bibinfo{author}{\bibfnamefont{C.}~\bibnamefont{Camilloni}},
  \bibnamefont{and} \bibinfo{author}{\bibfnamefont{G.}~\bibnamefont{Bussi}},
  \bibinfo{journal}{Comput. Phys. Comm.} \textbf{\bibinfo{volume}{185}},
  \bibinfo{pages}{604} (\bibinfo{year}{2014}).

\bibitem[{\citenamefont{Jensen et~al.}(2010)\citenamefont{Jensen, Thomsen, von
  Solms, Wierzchowski, Walsh, Koh, Sloan, Wu, and Sum}}]{2010-JPCB-Jensen-Sum}
\bibinfo{author}{\bibfnamefont{L.}~\bibnamefont{Jensen}},
  \bibinfo{author}{\bibfnamefont{K.}~\bibnamefont{Thomsen}},
  \bibinfo{author}{\bibfnamefont{N.}~\bibnamefont{von Solms}},
  \bibinfo{author}{\bibfnamefont{S.}~\bibnamefont{Wierzchowski}},
  \bibinfo{author}{\bibfnamefont{M.}~\bibnamefont{Walsh}},
  \bibinfo{author}{\bibfnamefont{C.}~\bibnamefont{Koh}},
  \bibinfo{author}{\bibfnamefont{E.}~\bibnamefont{Sloan}},
  \bibinfo{author}{\bibfnamefont{D.}~\bibnamefont{Wu}}, \bibnamefont{and}
  \bibinfo{author}{\bibfnamefont{A.}~\bibnamefont{Sum}}, \bibinfo{journal}{J.
  Phys. Chem. B} \textbf{\bibinfo{volume}{114}}, \bibinfo{pages}{5775}
  (\bibinfo{year}{2010}).

\bibitem[{\citenamefont{Bagherzadeh et~al.}(2015)\citenamefont{Bagherzadeh,
  Alavi, Ripmeester, and Englezos}}]{2015-PCCP-Alireza-Englezos}
\bibinfo{author}{\bibfnamefont{S.~A.} \bibnamefont{Bagherzadeh}},
  \bibinfo{author}{\bibfnamefont{S.}~\bibnamefont{Alavi}},
  \bibinfo{author}{\bibfnamefont{J.}~\bibnamefont{Ripmeester}},
  \bibnamefont{and} \bibinfo{author}{\bibfnamefont{P.}~\bibnamefont{Englezos}},
  \bibinfo{journal}{Phys. Chem. Chem. Phys.} \textbf{\bibinfo{volume}{17}},
  \bibinfo{pages}{9984} (\bibinfo{year}{2015}).

\bibitem[{\citenamefont{Conde and Vega}(2010)}]{2010-JCP-Conde-Vega}
\bibinfo{author}{\bibfnamefont{M.}~\bibnamefont{Conde}} \bibnamefont{and}
  \bibinfo{author}{\bibfnamefont{C.}~\bibnamefont{Vega}}, \bibinfo{journal}{J.
  Chem. Phys.} \textbf{\bibinfo{volume}{133}}, \bibinfo{pages}{064507}
  (\bibinfo{year}{2010}).

\bibitem[{\citenamefont{Martin and Siepmann}(1998)}]{1998-JPCB-Martin-Siepmann}
\bibinfo{author}{\bibfnamefont{M.}~\bibnamefont{Martin}} \bibnamefont{and}
  \bibinfo{author}{\bibfnamefont{J.}~\bibnamefont{Siepmann}},
  \bibinfo{journal}{J. Phys. Chem. B} \textbf{\bibinfo{volume}{102}},
  \bibinfo{pages}{2569} (\bibinfo{year}{1998}).

\bibitem[{\citenamefont{Wang et~al.}(2004)\citenamefont{Wang, Wolf, Caldwell,
  Kollman, and Case}}]{2004-JCC-Wang-Case}
\bibinfo{author}{\bibfnamefont{J.}~\bibnamefont{Wang}},
  \bibinfo{author}{\bibfnamefont{R.}~\bibnamefont{Wolf}},
  \bibinfo{author}{\bibfnamefont{J.}~\bibnamefont{Caldwell}},
  \bibinfo{author}{\bibfnamefont{P.}~\bibnamefont{Kollman}}, \bibnamefont{and}
  \bibinfo{author}{\bibfnamefont{D.}~\bibnamefont{Case}}, \bibinfo{journal}{J.
  Comput. Chem.} \textbf{\bibinfo{volume}{25}}, \bibinfo{pages}{1157}
  (\bibinfo{year}{2004}).

\bibitem[{\citenamefont{Case et~al.}(2014)\citenamefont{Case, Berryman, betz,
  Cai, Cerutti, Cheatham, Darden, Duke, Gohlke, Goetz et~al.}}]{AMBER14}
\bibinfo{author}{\bibfnamefont{D.}~\bibnamefont{Case}},
  \bibinfo{author}{\bibfnamefont{J.}~\bibnamefont{Berryman}},
  \bibinfo{author}{\bibfnamefont{R.}~\bibnamefont{betz}},
  \bibinfo{author}{\bibfnamefont{Q.}~\bibnamefont{Cai}},
  \bibinfo{author}{\bibfnamefont{D.}~\bibnamefont{Cerutti}},
  \bibinfo{author}{\bibfnamefont{T.}~\bibnamefont{Cheatham}},
  \bibinfo{author}{\bibfnamefont{T.}~\bibnamefont{Darden}},
  \bibinfo{author}{\bibfnamefont{R.}~\bibnamefont{Duke}},
  \bibinfo{author}{\bibfnamefont{H.}~\bibnamefont{Gohlke}},
  \bibinfo{author}{\bibfnamefont{A.}~\bibnamefont{Goetz}},
  \bibnamefont{et~al.}, \emph{\bibinfo{title}{AMBER 14}}
  (\bibinfo{publisher}{University of California: San Francisco, California},
  \bibinfo{year}{2014}).

\bibitem[{\citenamefont{Smith and Dang}(1994)}]{1994-JCP-Smith-Dang}
\bibinfo{author}{\bibfnamefont{D.}~\bibnamefont{Smith}} \bibnamefont{and}
  \bibinfo{author}{\bibfnamefont{L.}~\bibnamefont{Dang}}, \bibinfo{journal}{J.
  Chem. Phys.} \textbf{\bibinfo{volume}{100}}, \bibinfo{pages}{3757}
  (\bibinfo{year}{1994}).

\bibitem[{\citenamefont{Lorentz}(1881)}]{1881-AP-Lorentz}
\bibinfo{author}{\bibfnamefont{H.}~\bibnamefont{Lorentz}},
  \bibinfo{journal}{Ann. Phys.} \textbf{\bibinfo{volume}{248}},
  \bibinfo{pages}{127} (\bibinfo{year}{1881}).

\bibitem[{\citenamefont{Berthelot}(1898)}]{1898-CR-Berthelot}
\bibinfo{author}{\bibfnamefont{D.}~\bibnamefont{Berthelot}},
  \bibinfo{journal}{Compt. Rendus} \textbf{\bibinfo{volume}{126}},
  \bibinfo{pages}{1703} (\bibinfo{year}{1898}).

\bibitem[{\citenamefont{Darden et~al.}(1993)\citenamefont{Darden, York, and
  Pedersen}}]{1993-JCP-Darden-Pedersen}
\bibinfo{author}{\bibfnamefont{T.}~\bibnamefont{Darden}},
  \bibinfo{author}{\bibfnamefont{D.}~\bibnamefont{York}}, \bibnamefont{and}
  \bibinfo{author}{\bibfnamefont{L.}~\bibnamefont{Pedersen}},
  \bibinfo{journal}{J. Chem. Phys.} \textbf{\bibinfo{volume}{98}},
  \bibinfo{pages}{10089} (\bibinfo{year}{1993}).

\bibitem[{\citenamefont{Essmann et~al.}(1995)\citenamefont{Essmann, Perera, and
  Berkowitz}}]{1995-JCP-Essmann-Berkowitz}
\bibinfo{author}{\bibfnamefont{U.}~\bibnamefont{Essmann}},
  \bibinfo{author}{\bibfnamefont{L.}~\bibnamefont{Perera}}, \bibnamefont{and}
  \bibinfo{author}{\bibfnamefont{M.}~\bibnamefont{Berkowitz}},
  \bibinfo{journal}{J. Chem. Phys.} \textbf{\bibinfo{volume}{103}},
  \bibinfo{pages}{8577} (\bibinfo{year}{1995}).

\bibitem[{\citenamefont{Kawata and
  Nagashima}(2001)}]{2001-CPL-Kawata-Nagashima}
\bibinfo{author}{\bibfnamefont{M.}~\bibnamefont{Kawata}} \bibnamefont{and}
  \bibinfo{author}{\bibfnamefont{U.}~\bibnamefont{Nagashima}},
  \bibinfo{journal}{Chem. Phys. Lett.} \textbf{\bibinfo{volume}{340}},
  \bibinfo{pages}{165} (\bibinfo{year}{2001}).

\bibitem[{\citenamefont{Takeuchi et~al.}(2013)\citenamefont{Takeuchi,
  Hiratsuka, Ohmura, Alavi, Sum, and Yasuoka}}]{2013-JCP-Takeuchi-Yasuoka}
\bibinfo{author}{\bibfnamefont{F.}~\bibnamefont{Takeuchi}},
  \bibinfo{author}{\bibfnamefont{M.}~\bibnamefont{Hiratsuka}},
  \bibinfo{author}{\bibfnamefont{R.}~\bibnamefont{Ohmura}},
  \bibinfo{author}{\bibfnamefont{S.}~\bibnamefont{Alavi}},
  \bibinfo{author}{\bibfnamefont{A.}~\bibnamefont{Sum}}, \bibnamefont{and}
  \bibinfo{author}{\bibfnamefont{K.}~\bibnamefont{Yasuoka}},
  \bibinfo{journal}{J. Chem. Phys.} \textbf{\bibinfo{volume}{138}},
  \bibinfo{pages}{124504} (\bibinfo{year}{2013}).

\bibitem[{\citenamefont{Berendsen et~al.}(1984)\citenamefont{Berendsen, Postma,
  van Gunsteren, DiNola, and Haak}}]{1984-JCP-Berendsen-Haak}
\bibinfo{author}{\bibfnamefont{H.}~\bibnamefont{Berendsen}},
  \bibinfo{author}{\bibfnamefont{J.}~\bibnamefont{Postma}},
  \bibinfo{author}{\bibfnamefont{W.}~\bibnamefont{van Gunsteren}},
  \bibinfo{author}{\bibfnamefont{A.}~\bibnamefont{DiNola}}, \bibnamefont{and}
  \bibinfo{author}{\bibfnamefont{J.~R.} \bibnamefont{Haak}},
  \bibinfo{journal}{J. Chem. Phys.} \textbf{\bibinfo{volume}{81}},
  \bibinfo{pages}{3684} (\bibinfo{year}{1984}).

\bibitem[{\citenamefont{Pronk et~al.}(2013)\citenamefont{Pronk, S.P\'all,
  Schulz, Larsson, Bjelkmar, Apostolov, Shirts, Smith, Kasson, van~der Spoel
  et~al.}}]{2013-Bioinformatics-Pronk-Lindakl}
\bibinfo{author}{\bibfnamefont{S.}~\bibnamefont{Pronk}},
  \bibinfo{author}{\bibnamefont{S.P\'all}},
  \bibinfo{author}{\bibfnamefont{R.}~\bibnamefont{Schulz}},
  \bibinfo{author}{\bibfnamefont{P.}~\bibnamefont{Larsson}},
  \bibinfo{author}{\bibfnamefont{P.}~\bibnamefont{Bjelkmar}},
  \bibinfo{author}{\bibfnamefont{R.}~\bibnamefont{Apostolov}},
  \bibinfo{author}{\bibfnamefont{M.}~\bibnamefont{Shirts}},
  \bibinfo{author}{\bibfnamefont{J.}~\bibnamefont{Smith}},
  \bibinfo{author}{\bibfnamefont{P.}~\bibnamefont{Kasson}},
  \bibinfo{author}{\bibfnamefont{D.}~\bibnamefont{van~der Spoel}},
  \bibnamefont{et~al.}, \bibinfo{journal}{Bioinformatics}
  \textbf{\bibinfo{volume}{29}}, \bibinfo{pages}{845} (\bibinfo{year}{2013}).

\bibitem[{\citenamefont{Evans and Holian}(1985)}]{1985-JCP-Evans-Holian}
\bibinfo{author}{\bibfnamefont{D.}~\bibnamefont{Evans}} \bibnamefont{and}
  \bibinfo{author}{\bibfnamefont{B.}~\bibnamefont{Holian}},
  \bibinfo{journal}{J. Chem. Phys.} \textbf{\bibinfo{volume}{83}},
  \bibinfo{pages}{4069} (\bibinfo{year}{1985}).

\bibitem[{\citenamefont{Parrinello and
  Rahman}(1981)}]{1981-JAP-Parrinello-Rahman}
\bibinfo{author}{\bibfnamefont{M.}~\bibnamefont{Parrinello}} \bibnamefont{and}
  \bibinfo{author}{\bibfnamefont{A.}~\bibnamefont{Rahman}},
  \bibinfo{journal}{J. Appl. Phys.} \textbf{\bibinfo{volume}{52}},
  \bibinfo{pages}{7182} (\bibinfo{year}{1981}).

\bibitem[{\citenamefont{Bussi et~al.}(2007)\citenamefont{Bussi, Donadio, and
  Parrinello}}]{2007-JCP-Bussi-Parrinello}
\bibinfo{author}{\bibfnamefont{G.}~\bibnamefont{Bussi}},
  \bibinfo{author}{\bibfnamefont{D.}~\bibnamefont{Donadio}}, \bibnamefont{and}
  \bibinfo{author}{\bibfnamefont{M.}~\bibnamefont{Parrinello}},
  \bibinfo{journal}{J. Chem. Phys.} \textbf{\bibinfo{volume}{126}},
  \bibinfo{pages}{014101} (\bibinfo{year}{2007}).

\bibitem[{\citenamefont{McCammon}(2006)}]{2006-CR-Adcock-McCammon}
\bibinfo{author}{\bibfnamefont{S.~A.~J.} \bibnamefont{McCammon}},
  \bibinfo{journal}{Chem. Rev.} \textbf{\bibinfo{volume}{106}},
  \bibinfo{pages}{1589} (\bibinfo{year}{2006}).

\bibitem[{\citenamefont{Spiwok et~al.}(2015)\citenamefont{Spiwok, Sucur, and
  Hosek}}]{2015-BA-Spiwok-Hosek}
\bibinfo{author}{\bibfnamefont{V.}~\bibnamefont{Spiwok}},
  \bibinfo{author}{\bibfnamefont{Z.}~\bibnamefont{Sucur}}, \bibnamefont{and}
  \bibinfo{author}{\bibfnamefont{P.}~\bibnamefont{Hosek}},
  \bibinfo{journal}{Biotechnology Adv.} \textbf{\bibinfo{volume}{33}},
  \bibinfo{pages}{1130} (\bibinfo{year}{2015}).

\bibitem[{\citenamefont{Bernardi et~al.}(2015)\citenamefont{Bernardi, Melo, and
  Schulten}}]{2015-BBA-Bernardi-Schulten}
\bibinfo{author}{\bibfnamefont{R.}~\bibnamefont{Bernardi}},
  \bibinfo{author}{\bibfnamefont{M.}~\bibnamefont{Melo}}, \bibnamefont{and}
  \bibinfo{author}{\bibfnamefont{K.}~\bibnamefont{Schulten}},
  \bibinfo{journal}{Biochim. Biophys. Acta.} \textbf{\bibinfo{volume}{1850}},
  \bibinfo{pages}{872} (\bibinfo{year}{2015}).

\bibitem[{\citenamefont{Maximova et~al.}(2016)\citenamefont{Maximova, Moffatt,
  Nussinov, and Shehu}}]{2016-Plos-Maximova-Shehu}
\bibinfo{author}{\bibfnamefont{T.}~\bibnamefont{Maximova}},
  \bibinfo{author}{\bibfnamefont{R.}~\bibnamefont{Moffatt}},
  \bibinfo{author}{\bibfnamefont{R.}~\bibnamefont{Nussinov}}, \bibnamefont{and}
  \bibinfo{author}{\bibfnamefont{A.}~\bibnamefont{Shehu}},
  \bibinfo{journal}{PLoS Comput. Biol.} \textbf{\bibinfo{volume}{12}},
  \bibinfo{pages}{e1004619} (\bibinfo{year}{2016}).

\bibitem[{\citenamefont{Pietrucci}(2017)}]{2017-RP-Pietrucci}
\bibinfo{author}{\bibfnamefont{F.}~\bibnamefont{Pietrucci}},
  \bibinfo{journal}{Rev. Phys.} \textbf{\bibinfo{volume}{2}},
  \bibinfo{pages}{32} (\bibinfo{year}{2017}).

\bibitem[{\citenamefont{Laio and Gervasio}(2008)}]{2008-RPP-Laio-Gervasio}
\bibinfo{author}{\bibfnamefont{A.}~\bibnamefont{Laio}} \bibnamefont{and}
  \bibinfo{author}{\bibfnamefont{F.}~\bibnamefont{Gervasio}},
  \bibinfo{journal}{Rep. Prog. Phys.} \textbf{\bibinfo{volume}{71}},
  \bibinfo{pages}{126601} (\bibinfo{year}{2008}).

\bibitem[{\citenamefont{Barducci et~al.}(2011)\citenamefont{Barducci, Bonomi,
  and Parrinello}}]{2011-CMS-Barducci-Parrinello}
\bibinfo{author}{\bibfnamefont{A.}~\bibnamefont{Barducci}},
  \bibinfo{author}{\bibfnamefont{M.}~\bibnamefont{Bonomi}}, \bibnamefont{and}
  \bibinfo{author}{\bibfnamefont{M.}~\bibnamefont{Parrinello}},
  \bibinfo{journal}{WIREs Comput. Mol. Sci.} \textbf{\bibinfo{volume}{1}},
  \bibinfo{pages}{826} (\bibinfo{year}{2011}).

\bibitem[{\citenamefont{Sutto et~al.}(2012)\citenamefont{Sutto, Marsili, and
  Gervasio}}]{2012-CMS-Sutto-Gervasio}
\bibinfo{author}{\bibfnamefont{L.}~\bibnamefont{Sutto}},
  \bibinfo{author}{\bibfnamefont{S.}~\bibnamefont{Marsili}}, \bibnamefont{and}
  \bibinfo{author}{\bibfnamefont{F.}~\bibnamefont{Gervasio}},
  \bibinfo{journal}{WIREs Comput. Mol. Sci.} \textbf{\bibinfo{volume}{2}},
  \bibinfo{pages}{771} (\bibinfo{year}{2012}).

\bibitem[{\citenamefont{Kastner}(2011)}]{2011-CMS-Kastner}
\bibinfo{author}{\bibfnamefont{J.}~\bibnamefont{Kastner}},
  \bibinfo{journal}{WIREs Comput. Mol. Sci.} \textbf{\bibinfo{volume}{1}},
  \bibinfo{pages}{932} (\bibinfo{year}{2011}).

\bibitem[{\citenamefont{Barducci et~al.}(2006)\citenamefont{Barducci, Chelli,
  Procacci, Schettino, Gervasio, and
  Parrinello}}]{2006-JACS-Barducci-Parrinello}
\bibinfo{author}{\bibfnamefont{A.}~\bibnamefont{Barducci}},
  \bibinfo{author}{\bibfnamefont{R.}~\bibnamefont{Chelli}},
  \bibinfo{author}{\bibfnamefont{P.}~\bibnamefont{Procacci}},
  \bibinfo{author}{\bibfnamefont{V.}~\bibnamefont{Schettino}},
  \bibinfo{author}{\bibfnamefont{F.}~\bibnamefont{Gervasio}}, \bibnamefont{and}
  \bibinfo{author}{\bibfnamefont{M.}~\bibnamefont{Parrinello}},
  \bibinfo{journal}{J. Am. Chem. Soc.} \textbf{\bibinfo{volume}{128}},
  \bibinfo{pages}{2705} (\bibinfo{year}{2006}).

\bibitem[{\citenamefont{Sutto and Gervasio}(2013)}]{2013-PNAS-Sutto-Gervasio}
\bibinfo{author}{\bibfnamefont{L.}~\bibnamefont{Sutto}} \bibnamefont{and}
  \bibinfo{author}{\bibfnamefont{F.}~\bibnamefont{Gervasio}},
  \bibinfo{journal}{Proc. Nat. Acad. Soc.} \textbf{\bibinfo{volume}{110}},
  \bibinfo{pages}{10616} (\bibinfo{year}{2013}).

\bibitem[{\citenamefont{Sicard and Senet}(2013)}]{2013-JCP-Sicard-Senet}
\bibinfo{author}{\bibfnamefont{F.}~\bibnamefont{Sicard}} \bibnamefont{and}
  \bibinfo{author}{\bibfnamefont{P.}~\bibnamefont{Senet}}, \bibinfo{journal}{J.
  Chem. Phys.} \textbf{\bibinfo{volume}{138}}, \bibinfo{pages}{235101}
  (\bibinfo{year}{2013}).

\bibitem[{\citenamefont{Giberti et~al.}(2015)\citenamefont{Giberti,
  Salvalaglio, and Parrinello}}]{2015-IUCrJ-Giberti-Parrinello}
\bibinfo{author}{\bibfnamefont{F.}~\bibnamefont{Giberti}},
  \bibinfo{author}{\bibfnamefont{M.}~\bibnamefont{Salvalaglio}},
  \bibnamefont{and}
  \bibinfo{author}{\bibfnamefont{M.}~\bibnamefont{Parrinello}},
  \bibinfo{journal}{IUCrJ} \textbf{\bibinfo{volume}{2}}, \bibinfo{pages}{256}
  (\bibinfo{year}{2015}).

\bibitem[{\citenamefont{Salvalaglio
  et~al.}(2015{\natexlab{a}})\citenamefont{Salvalaglio, Mazzotti, and
  Parrinello}}]{2015-FD-Salvalaglio-Parrinello}
\bibinfo{author}{\bibfnamefont{M.}~\bibnamefont{Salvalaglio}},
  \bibinfo{author}{\bibfnamefont{M.}~\bibnamefont{Mazzotti}}, \bibnamefont{and}
  \bibinfo{author}{\bibfnamefont{M.}~\bibnamefont{Parrinello}},
  \bibinfo{journal}{Faraday Discus.} \textbf{\bibinfo{volume}{179}},
  \bibinfo{pages}{291} (\bibinfo{year}{2015}{\natexlab{a}}).

\bibitem[{\citenamefont{Gimondi and
  Salvalaglio}(2017)}]{2017-JCP-Gimondi-Salvalaglio}
\bibinfo{author}{\bibfnamefont{I.}~\bibnamefont{Gimondi}} \bibnamefont{and}
  \bibinfo{author}{\bibfnamefont{M.}~\bibnamefont{Salvalaglio}},
  \bibinfo{journal}{J. Chem. Phys.} \textbf{\bibinfo{volume}{147}},
  \bibinfo{pages}{114502} (\bibinfo{year}{2017}).

\bibitem[{\citenamefont{Barducci et~al.}(2008)\citenamefont{Barducci, Bussi,
  and Parrinello}}]{2008-PRL-Barducci-Parrinello}
\bibinfo{author}{\bibfnamefont{A.}~\bibnamefont{Barducci}},
  \bibinfo{author}{\bibfnamefont{G.}~\bibnamefont{Bussi}}, \bibnamefont{and}
  \bibinfo{author}{\bibfnamefont{M.}~\bibnamefont{Parrinello}},
  \bibinfo{journal}{Phys. Rev. Lett.} \textbf{\bibinfo{volume}{100}},
  \bibinfo{pages}{020603} (\bibinfo{year}{2008}).

\bibitem[{\citenamefont{Dama et~al.}(2014)\citenamefont{Dama, Parrinello, and
  Voth}}]{2014-PRL-Dama-Voth}
\bibinfo{author}{\bibfnamefont{J.}~\bibnamefont{Dama}},
  \bibinfo{author}{\bibfnamefont{M.}~\bibnamefont{Parrinello}},
  \bibnamefont{and} \bibinfo{author}{\bibfnamefont{G.}~\bibnamefont{Voth}},
  \bibinfo{journal}{Phys. Rev. Lett.} \textbf{\bibinfo{volume}{112}},
  \bibinfo{pages}{240602} (\bibinfo{year}{2014}).

\bibitem[{\citenamefont{Zhang and Voth}(2011)}]{2011-JCTC-Zhang-Voth}
\bibinfo{author}{\bibfnamefont{Y.}~\bibnamefont{Zhang}} \bibnamefont{and}
  \bibinfo{author}{\bibfnamefont{G.}~\bibnamefont{Voth}}, \bibinfo{journal}{J.
  Chem. Theory Comput.} \textbf{\bibinfo{volume}{7}}, \bibinfo{pages}{2277}
  (\bibinfo{year}{2011}).

\bibitem[{\citenamefont{Nishizawa and
  Nishizawa}(2013)}]{2013-BJ-Nishizawa-Nishizawa}
\bibinfo{author}{\bibfnamefont{M.}~\bibnamefont{Nishizawa}} \bibnamefont{and}
  \bibinfo{author}{\bibfnamefont{K.}~\bibnamefont{Nishizawa}},
  \bibinfo{journal}{Biophys. J.} \textbf{\bibinfo{volume}{104}},
  \bibinfo{pages}{1038} (\bibinfo{year}{2013}).

\bibitem[{\citenamefont{Paci and Karplus}(1999)}]{1999-JMB-Paci-Karplus}
\bibinfo{author}{\bibfnamefont{E.}~\bibnamefont{Paci}} \bibnamefont{and}
  \bibinfo{author}{\bibfnamefont{M.}~\bibnamefont{Karplus}},
  \bibinfo{journal}{J. Mol. Biol.} \textbf{\bibinfo{volume}{288}},
  \bibinfo{pages}{441} (\bibinfo{year}{1999}).

\bibitem[{\citenamefont{Marchi and Ballone}(1999)}]{1999-JCP-Marchi-Ballone}
\bibinfo{author}{\bibfnamefont{M.}~\bibnamefont{Marchi}} \bibnamefont{and}
  \bibinfo{author}{\bibfnamefont{P.}~\bibnamefont{Ballone}},
  \bibinfo{journal}{J. Chem. Phys.} \textbf{\bibinfo{volume}{110}},
  \bibinfo{pages}{3697} (\bibinfo{year}{1999}).

\bibitem[{\citenamefont{Camilloni et~al.}(2011)\citenamefont{Camilloni,
  Broglia, and Tiana}}]{2011-JCP-Camilloni-Tiana}
\bibinfo{author}{\bibfnamefont{C.}~\bibnamefont{Camilloni}},
  \bibinfo{author}{\bibfnamefont{R.}~\bibnamefont{Broglia}}, \bibnamefont{and}
  \bibinfo{author}{\bibfnamefont{G.}~\bibnamefont{Tiana}}, \bibinfo{journal}{J.
  Chem. Phys.} \textbf{\bibinfo{volume}{134}}, \bibinfo{pages}{045105}
  (\bibinfo{year}{2011}).

\bibitem[{\citenamefont{Sicard and Striolo}(2016)}]{2016-FD-Sicard-Striolo}
\bibinfo{author}{\bibfnamefont{F.}~\bibnamefont{Sicard}} \bibnamefont{and}
  \bibinfo{author}{\bibfnamefont{A.}~\bibnamefont{Striolo}},
  \bibinfo{journal}{Faraday Discuss.} \textbf{\bibinfo{volume}{191}},
  \bibinfo{pages}{287} (\bibinfo{year}{2016}).

\bibitem[{\citenamefont{Allen et~al.}(2004)\citenamefont{Allen, Andersen, and
  Roux}}]{2004-PNAS-Allen-Roux}
\bibinfo{author}{\bibfnamefont{T.}~\bibnamefont{Allen}},
  \bibinfo{author}{\bibfnamefont{O.}~\bibnamefont{Andersen}}, \bibnamefont{and}
  \bibinfo{author}{\bibfnamefont{B.}~\bibnamefont{Roux}},
  \bibinfo{journal}{Proc. Nat. Acad. Sci. U.S.A.}
  \textbf{\bibinfo{volume}{101}}, \bibinfo{pages}{117} (\bibinfo{year}{2004}).

\bibitem[{\citenamefont{Zhu and Hummer}(2012)}]{2012-JCTC-Zhu-Hummer}
\bibinfo{author}{\bibfnamefont{F.}~\bibnamefont{Zhu}} \bibnamefont{and}
  \bibinfo{author}{\bibfnamefont{G.}~\bibnamefont{Hummer}},
  \bibinfo{journal}{J. Chem. Theory Comput.} \textbf{\bibinfo{volume}{8}},
  \bibinfo{pages}{3759} (\bibinfo{year}{2012}).

\bibitem[{\citenamefont{Grossfiled}(2013)}]{WHAM}
\bibinfo{author}{\bibfnamefont{A.}~\bibnamefont{Grossfiled}},
  \bibinfo{journal}{http://membrane.urmc.rochester.edu/content/wham}
  (\bibinfo{year}{2013}).

\bibitem[{\citenamefont{Schneider and
  Reuter}(2014)}]{2014-JPCL-Schnelder-Reuter}
\bibinfo{author}{\bibfnamefont{J.}~\bibnamefont{Schneider}} \bibnamefont{and}
  \bibinfo{author}{\bibfnamefont{K.}~\bibnamefont{Reuter}},
  \bibinfo{journal}{J. Phys. Chem. Lett.} \textbf{\bibinfo{volume}{5}},
  \bibinfo{pages}{3859} (\bibinfo{year}{2014}).

\bibitem[{\citenamefont{Salvalaglio
  et~al.}(2015{\natexlab{b}})\citenamefont{Salvalaglio, Perego, F.Giberti,
  Mazzotti, and Parrinello}}]{2015-PNAS-Salvalaglio-Parrinello}
\bibinfo{author}{\bibfnamefont{M.}~\bibnamefont{Salvalaglio}},
  \bibinfo{author}{\bibfnamefont{C.}~\bibnamefont{Perego}},
  \bibinfo{author}{\bibnamefont{F.Giberti}},
  \bibinfo{author}{\bibfnamefont{M.}~\bibnamefont{Mazzotti}}, \bibnamefont{and}
  \bibinfo{author}{\bibfnamefont{M.}~\bibnamefont{Parrinello}},
  \bibinfo{journal}{Proc. Nat. Acad. Sci. U.S.A.}
  \textbf{\bibinfo{volume}{112}}, \bibinfo{pages}{E6}
  (\bibinfo{year}{2015}{\natexlab{b}}).

\bibitem[{\citenamefont{Tiwary et~al.}(2015{\natexlab{b}})\citenamefont{Tiwary,
  Limongelli, Salvalaglio, and Parrinello}}]{2015-PNAS-Tiwary-Parrinello}
\bibinfo{author}{\bibfnamefont{P.}~\bibnamefont{Tiwary}},
  \bibinfo{author}{\bibfnamefont{V.}~\bibnamefont{Limongelli}},
  \bibinfo{author}{\bibfnamefont{M.}~\bibnamefont{Salvalaglio}},
  \bibnamefont{and}
  \bibinfo{author}{\bibfnamefont{M.}~\bibnamefont{Parrinello}},
  \bibinfo{journal}{Proc. Nat. Acad. Sci. U.S.A.}
  \textbf{\bibinfo{volume}{112}}, \bibinfo{pages}{E386}
  (\bibinfo{year}{2015}{\natexlab{b}}).

\bibitem[{\citenamefont{Piaggi et~al.}(2016)\citenamefont{Piaggi, Valssonbc,
  and Parrinello}}]{2016-FD-Piaggi-Parrinello}
\bibinfo{author}{\bibfnamefont{P.}~\bibnamefont{Piaggi}},
  \bibinfo{author}{\bibfnamefont{O.}~\bibnamefont{Valssonbc}},
  \bibnamefont{and}
  \bibinfo{author}{\bibfnamefont{M.}~\bibnamefont{Parrinello}},
  \bibinfo{journal}{Faraday Discuss.} \textbf{\bibinfo{volume}{195}},
  \bibinfo{pages}{557} (\bibinfo{year}{2016}).

\bibitem[{\citenamefont{Bochicchio et~al.}(2017)\citenamefont{Bochicchio,
  Salvalaglio, and Pavan}}]{2017-NC-Bochicchio-Pavan}
\bibinfo{author}{\bibfnamefont{D.}~\bibnamefont{Bochicchio}},
  \bibinfo{author}{\bibfnamefont{M.}~\bibnamefont{Salvalaglio}},
  \bibnamefont{and} \bibinfo{author}{\bibfnamefont{G.}~\bibnamefont{Pavan}},
  \bibinfo{journal}{Nat. Commun.} \textbf{\bibinfo{volume}{8}},
  \bibinfo{pages}{147} (\bibinfo{year}{2017}).

\end{thebibliography}


\begin{thebibliography}{99}
%
\bibitem{2008-PRL-Barducci-Parrinello-SI} A. Barducci, G. Bussi and M. Parrinello, Phys. Rev. Lett. (2008), \textbf{100}, 020603.
\bibitem{2014-PRL-Dama-Voth-SI} J. Dama, M. Parrinello and G. Voth, Phys. Rev. Lett. (2014), \textbf{112}, 240602.
\bibitem{2008-RPP-Laio-Gervasio-SI} A. Laio and F. Gervasio, Rep. Prog. Phys. (2008), \textbf{71}, 126601.
\bibitem{1977-JCP-Torrie-Valleau-SI} G. Torrie and J.P.Valleau, J. Comput. Phys. (1977), \textbf{23}, 187-199.
\bibitem{2011-CMS-Kastner-SI} J. Kastner, WIREs Comput. Mol. Sci. (2011), \textbf{1}, 932-942.
\bibitem{1999-JMB-Paci-Karplus-SI} E. Paci and M. Karplus, J. Mol. Biol. (1999), \textbf{288}, 441-459.
\bibitem{1999-JCP-Marchi-Ballone-SI} M. Marchi and P. Ballone, J. Chem. Phys. (1999), \textbf{110}, 3697-3702.
\bibitem{2011-JCP-Camilloni-Tiana-SI} C. Camilloni, R. Broglia and G. Tiana, J. Chem. Phys. (2011), \textbf{134}, 045105.
\bibitem{2016-FD-Sicard-Striolo-SI} F. Sicard and A. Striolo, Faraday Discuss. (2016), \textbf{191}, 287-304.
\bibitem{WHAM-SI} A. Grossfiled, WHAM: the weighted histogram analysis method, version 2.0.9. http://membrane.urmc.rochester.edu/content/wham.
\bibitem{2010-JCTC-Hub-vanderSpoel-SI} J. Hub, B. de Groot and D. van der Spoel, J. Chem. Theory Comput. (2010), \textbf{6}, 3713-3720.
\bibitem{1940-Physica-Kramers-SI} H. Kramers, Physica (1940), \textbf{7}, 284.
\bibitem{1990-RMP-Hanggi-Borkovec-SI} P. Hanggi, P. Talkner and M. Borkovec, Rev. Mod. Phys. (1990), \textbf{82}, 251-341.
\bibitem{2005-Chaos-Pollak-Talkner-SI} E. Pollak and P. Talkner, Chaos (2005), \textbf{15}, 026116.
\bibitem{2010-QRB-Zhou-SI} H.-X. Zhou, Quaterly Rev. Biophys. (2010), \textbf{43}, 219-293.
\bibitem{2017-ELSEVIER-Peters-SI} B. Peters, Reaction Rate Theory and Rare Events, 1st Ed.; Elsevier: Amsterdam, The Netherlands, 2017.
\bibitem{2018-arXiv-Sicard-SI} F. Sicard, arXiv:1803.03490 [cond-mat.stat-mech] (2018).
\bibitem{2013-PRL-Tiwary-Parrinello-SI} P. Tiwary and M. Parrinello, Phys. Rev. Lett. (2013), \textbf{111}, 230602.
\bibitem{2014-JCTC-Salvalaglio-Parrinello-SI} M. Salvalaglio, P. Tiwary and M. Parrinello, J. Chem. Theory Comput. (2014), \textbf{10}, 1420-1425.
\bibitem{cranR-SI} The Comprehensive R Archive Network. https://cran.r-project.org/ (accessed January 17, 2018)
%
\end{thebibliography}
%\bibliographystyle{rsc} %the RSC's .bst file

%%%%%%%%%% Merge with supplemental materials %%%%%%%%%%
\pagebreak
\widetext
\begin{center}
\textbf{\large Emergent Properties of Antiagglomerant Films Control Methane Transport: Implications for Hydrate Management}
\vskip 0.5cm
\textbf{\large Supporting Information}
\end{center}
%%%%%%%%%% Merge with supplemental materials %%%%%%%%%%
%%%%%%%%%% Prefix a "S" to all equations, figures, tables and reset the counter %%%%%%%%%%
\setcounter{equation}{0}
\setcounter{figure}{0}
\setcounter{table}{0}
\setcounter{page}{1}
\makeatletter
\renewcommand{\theequation}{S\arabic{equation}}
\renewcommand{\thefigure}{S\arabic{figure}}
\renewcommand{\bibnumfmt}[1]{[S#1]}
\renewcommand{\citenumfont}[1]{S#1}
%%%%%%%%%% Prefix a "S" to all equations, figures, tables and reset the counter %%%%%%%%%%
\section{Identification of representative pathways}
\textbf{Metadynamics simulations.} The well-tempered metadynamics (WT-metaD) 
framework~\cite{2008-PRL-Barducci-Parrinello-SI,2014-PRL-Dama-Voth-SI} was implemented to identify 
representative pathways available to the methane molecule to cross the antiagglomerants (AAs) film. 
We used the three Cartesian coordinates ($X$, $Y$ and $Z$) of the \textit{free} methane molecule 
as collective variables (CVs). A Gaussian potential was used as 
time-dependent bias, $V_{\textrm{bias}}(s,t)$~\cite{2008-RPP-Laio-Gervasio-SI}:
\begin{equation}
V_{\textrm{bias}}(s,t) = \omega \sum_{t'<t} \exp\Big[-\frac{(s(t)-s(t'))^2}{2 \sigma^2} \Big]~.
\label{metaD-potential-SI}
\end{equation}
In Eq.~\ref{metaD-potential-SI}, $\omega$ is the height of the biasing potential, $\sigma$ is the width, 
$t$ is the time, and $s$ is the collective variable. Following the algorithm introduced by 
Barducci et al.~\cite{2008-PRL-Barducci-Parrinello-SI}, a Gaussian-shaped potential was deposited every $\tau_G = 2$ ps, 
with height $\omega = \omega_0 e^{-V(s,t)/(f-1)T}$, where $\omega_0 = 5$ kJ/mol is the initial height, 
$T=277$ K is the temperature of the simulation, and $f\equiv(T+\Delta T)/T =25$ is the bias factor with $\Delta T$ 
a parameter with the dimension of a temperature. In the implementation, the resolution of the recovered 
free-energy (FE) surface is determined by the width of the Gaussian $\sigma=0.25$ nm along the $X$, $Y$ and $Z$ directions. 
WT-metaD simulations were run restraining the position of the AA layer, while allowing the hydrocarbon molecules, both at the interface and in the bulk phase, to move freely.
Soft walls were added on both sides of the interfacial layer to limit sampling inside the AA film.
\begin{figure}[h]
\includegraphics[width=0.45 \textwidth, angle=-0]{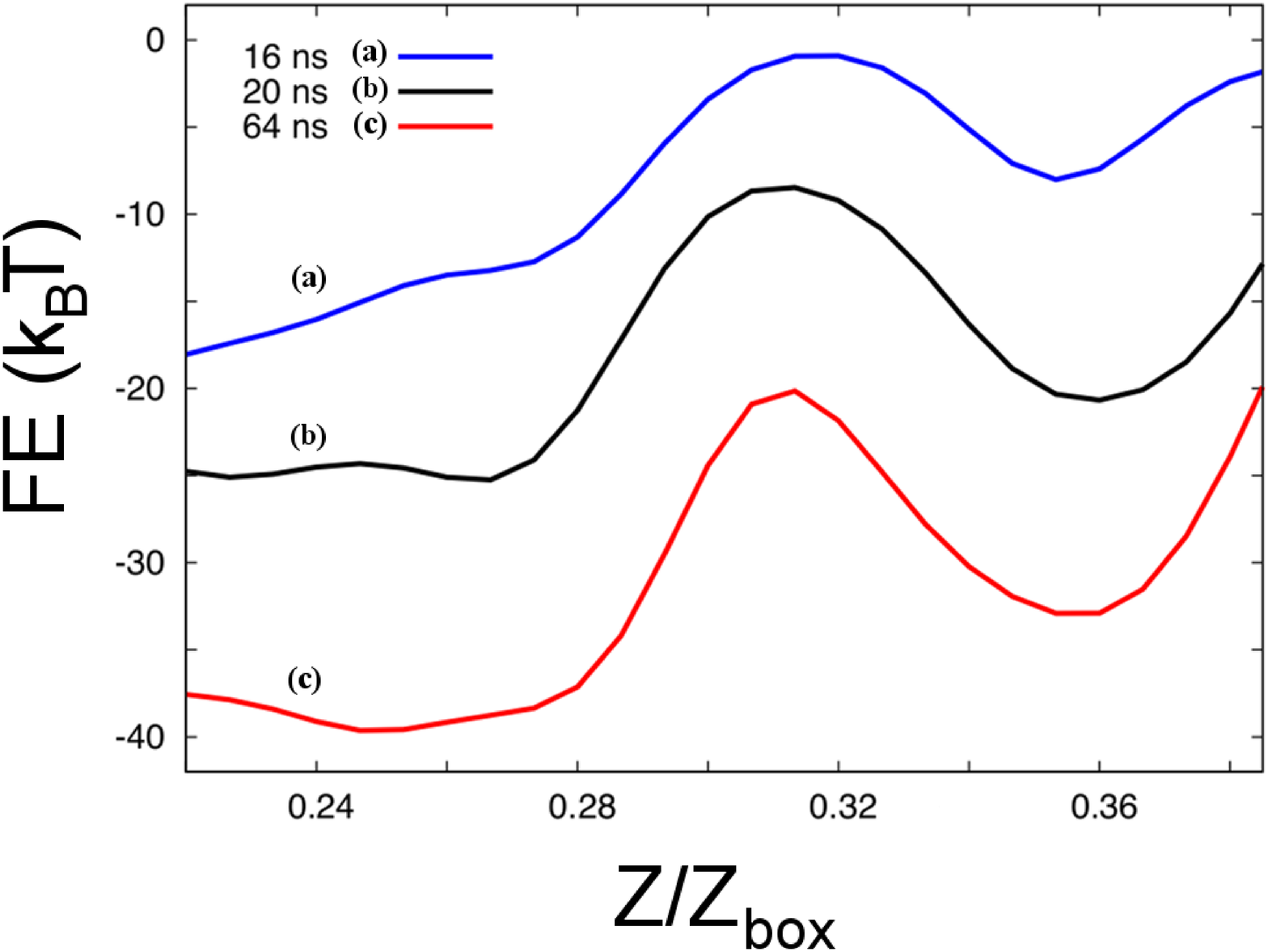}
 \caption{Reconstructed free-energy profile (FEP), along the \textit{minimal} pathway as a function of time. 
 The profile was reconstructed after $16$ ns (a), $20$ ns (b), and $64$ ns (c). 
 The $x$-axis corresponds to the $Z$-Cartesian coordinate of methane expressed in reduced units, 
 $Z/Z_{\textrm{box}}$, with $Z_{\textrm{box}}$ the size of the simulation box along the $Z$ direction. }
\label{figS1}
\end{figure}

Figure~\ref{figS1} shows the temporal evolution of the FEP reconstructed 
along the \textit{minimal} FE path, 
from the $170$ ns WT-metaD simulation. The FEP reached \textit{local} convergence 
after $20$ ns. However, after $64$ ns the FEP began to change. Analysis of the simulation 
confirmed that the change was a result of unphysical expulsion of hydrocarbon chain from the interfacial 
region caused by the too-high biased potential added.  
This indicated the need to account for the \textit{local} behaviour of the 
liquid hydrocarbon molecules. Indeed, the bias potential became sufficiently high to expel hydrocarbon 
molecules from the interfacial layer. The FEP reconstructed from the WT-metaD simulation shows the same 
qualitative behaviour as the one obtained within the umbrella sampling (US)/adiabatic biased molecular dynamics (ABMD) framework reported in the main text, with 
quantitative differences resulting from numerical artifacts.\\

\textbf{Umbrella sampling simulations.} Once the possible pathways across the AA film were identified, 
the potential of mean force (PMF) along them was rigorously calculated using US~\cite{1977-JCP-Torrie-Valleau-SI,2011-CMS-Kastner-SI}, with the $Z-$ Cartesian 
coordinate along the pathways as CV. The system was free in the $X-Y$ plane. To design the 
US windows, the ABMD 
framework was used~\cite{1999-JMB-Paci-Karplus-SI,1999-JCP-Marchi-Ballone-SI,
2011-JCP-Camilloni-Tiana-SI,2016-FD-Sicard-Striolo-SI}. To reconstruct the \textit{minimal} and \textit{intermediate} FEPs 
discussed in the main text, the authors generated $18$ and $33$ US windows, respectively, allowing sufficient overlap 
between adjacent windows. Upon completion of the US simulations, FEPs were calculated from the final $4$ ns 
of simulation time using the weighted histogram analysis method (WHAM)~\cite{WHAM-SI}. Statistical error analysis was 
conducted using the integrated Monte Carlo bootstrapping framework~\cite{2010-JCTC-Hub-vanderSpoel-SI}, 
implemented in WHAM, using 100 resampling trials.
\begin{figure}[h]
\includegraphics[width=0.45 \textwidth, angle=-0]{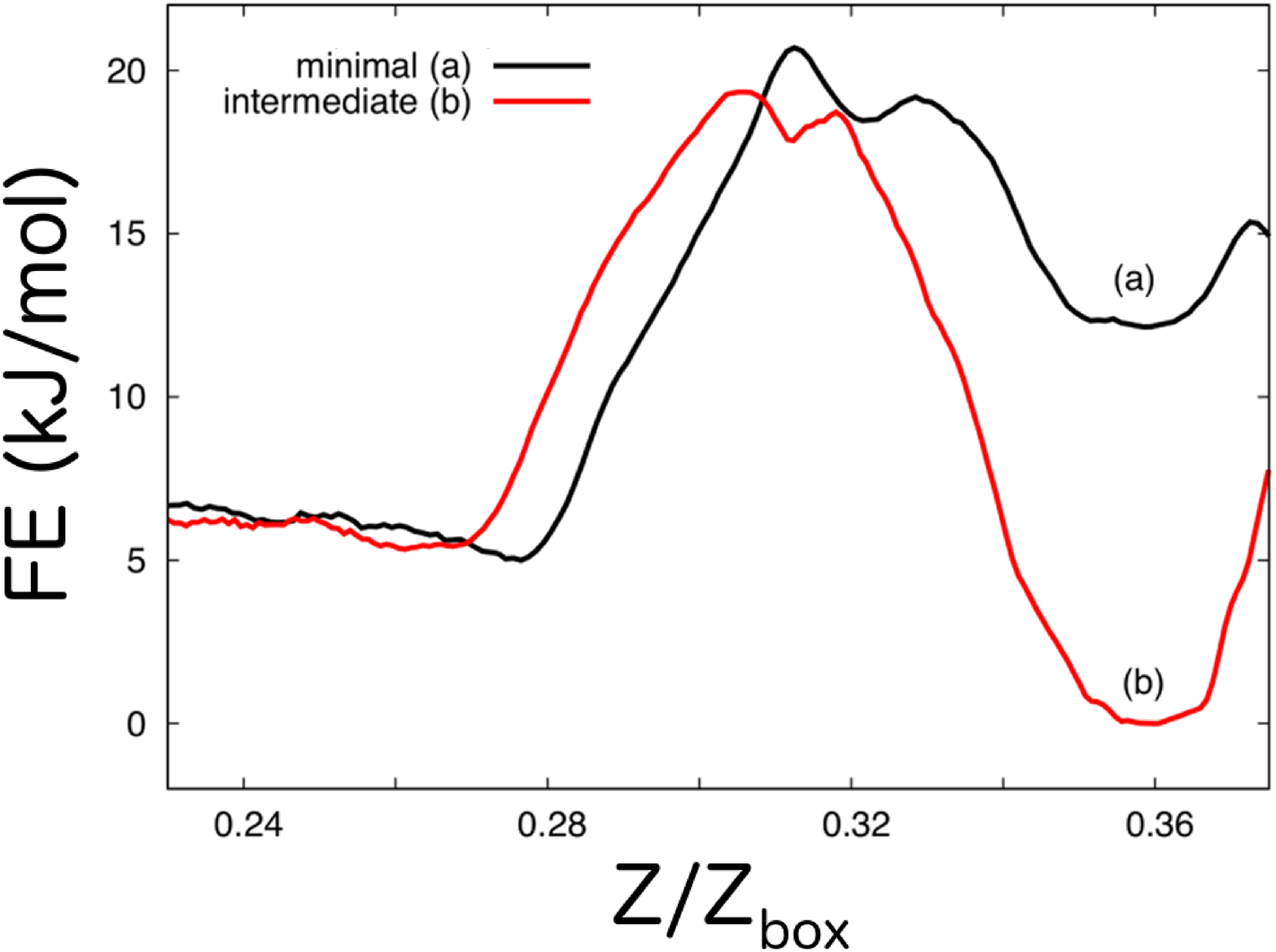}
 \caption{FEPs associated with the \textit{minimal} (a) and \textit{intermediate} (b) pathways, 
 obtained within the US/ABMD framework. The AA layer is flexible, although the central nitrogen atom of each AA 
 molecule remains fixed. The $x$-axis corresponds to the $Z$-Cartesian coordinate of methane expressed 
 in reduced units, $Z/Z_{\textrm{box}}$.
 }
\label{figS2}
\end{figure}

Figure~\ref{figS2} compares the FEPs discussed in the main text and obtained within US/ABMD frameworks. The activation energies 
of capture and escape are $\approx 16$ and $\approx 9$ kJ/mol, respectively, for the \textit{minimal} FEP, 
and $\approx 14.5$ and $\approx 20$ kJ/mol, respectively, for the \textit{intermediate} FEP.

\section{Dynamical analysis}
\begin{figure}[h]
\includegraphics[width=0.95 \textwidth, angle=-0]{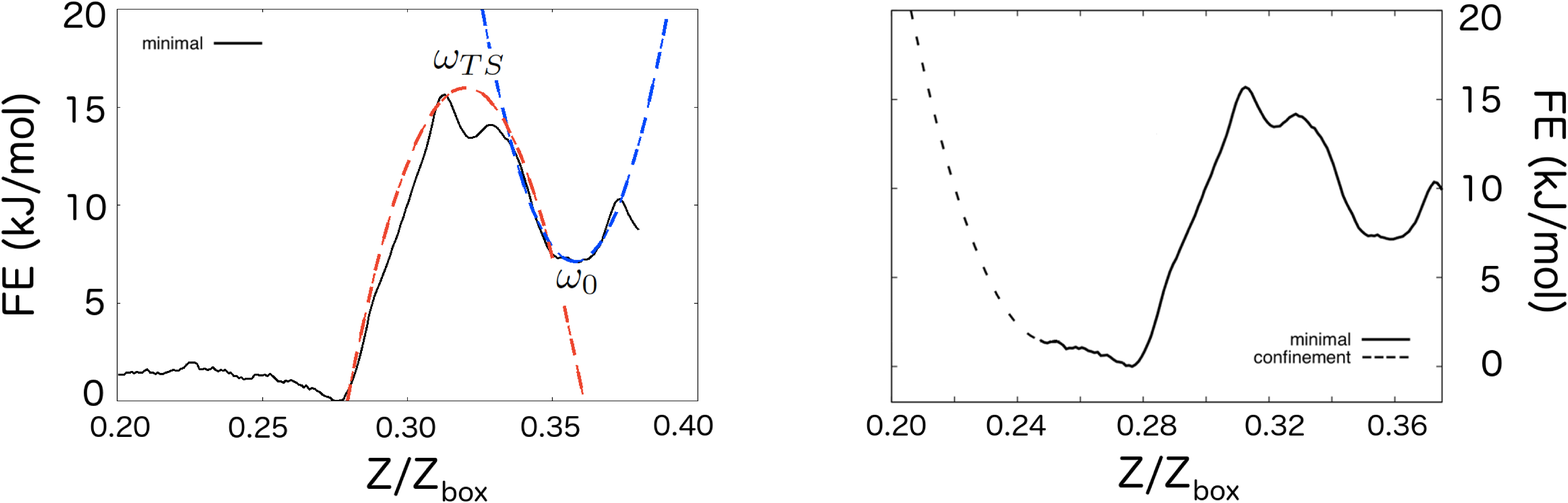}
 \caption{Fitting of the FEP within parabolic potentials to extract parameters 
 for the Kramers theory (left). Fitting yields   
$\omega_0 \approx 100 \times 10^{13}~s^{-1}$ and $\omega_{TS} \approx 230 \times 10^{13}~s^{-1}$. 
Representation of the external potential (black dashed line) used in the calculation of the characteristic 
time scale for methane capture (right).}
\label{figS3}
\end{figure}

\textbf{Kramers theory.} The authors estimated the transition rates along the FE pathways within 
the Kramers theory framework~\cite{1940-Physica-Kramers-SI,1990-RMP-Hanggi-Borkovec-SI,
2005-Chaos-Pollak-Talkner-SI,2010-QRB-Zhou-SI,2017-ELSEVIER-Peters-SI,2018-arXiv-Sicard-SI}. 
In the strong friction regime of interest here, the reaction rate, $k$, is given as~\cite{2017-ELSEVIER-Peters-SI}:
\begin{equation}
\label{Kramers-shortEQ-SI}
k = \frac{\omega_0 \omega_{TS}}{2\pi \gamma} e^{-\Delta F/k_B T} \, ,
\end{equation}
where $\gamma=6\pi\eta R/m$ is the friction coefficient, with $\eta$ the effective viscosity, 
and $\Delta F$ the height of the FE barrier. The parameters $\omega_0$ and $\omega_{TS}$ represent 
the stiffness of the potential well and the barrier, respectively, when modeled as parabolic potentials:
\begin{equation}
V(q) = V_{TS} - \frac{1}{2}m\omega_{TS}^2 (q-q_{TS})^2 \,.
\end{equation}
The nonlinear least-squares Marquardt-Levenberg algorithm was implemented to fit the parameters 
$\omega_0$ and $\omega_{TS}$, as shown in Figure~\ref{figS3}, left panel. We obtained  
$\omega_{TS} \approx 230 \times 10^{13}~s^{-1}$ and $\omega_0 \approx 100 \times 10^{13}~s^{-1}$ 
for the \textit{minimal} FE pathway, and $\omega_{TS} \approx 100 \times 10^{13}~s^{-1}$ 
and $\omega_0 \approx 130 \times 10^{13}~s^{-1}$ for the \textit{intermediate} FE pathway.
Figure~\ref{figS3}, right panel, shows an added external potential at $Z/Z_{\textrm{box}} \approx 0.25$, 
similar to the FE barrier observed for $Z/Z_{\textrm{box}} \in [0.27,0.31]$, to quantify the characteristic 
time scale for methane capture along the \textit{minimal} FE path.
\\

\textbf{Analysis of the transition times.} Building on the recent method of Parrinello, Salvalaglio, 
and Tiwary~\cite{2013-PRL-Tiwary-Parrinello-SI,2014-JCTC-Salvalaglio-Parrinello-SI}, we extended the metaD scope 
to assess numerically the characteristic time scales for methane escape and capture. We denote by $\tau$ 
the \textit{physical} mean transition time for the methane to pass over the energy barrier, 
and by $\tau_M$ the mean transition time obtained from the WT-metaD run. The latter is linked 
to the \textit{physical} mean transition time, $\tau$, by means of the acceleration factor
\begin{equation}
 \alpha(t) = \tau/\tau_M = \braket{e^{\beta V_{\textrm{bias}}(s,t)}}_M \, ,
\end{equation} 
where the angular brakets denote an average over a WT-metaD run confined to a metastable basin, 
and $V_{\textrm{bias}}(s,t)$ is the WT-metaD time-dependent bias defined in Eq.~\ref{metaD-potential-SI}. 
Because we are not interested in the diffusive converged limit of the metaD simulation, we increased the time lag between two successive Gaussian depositions in the WT-metaD framework, 
$\tau_G = 400$ ps~\cite{2013-PRL-Tiwary-Parrinello-SI,2014-JCTC-Salvalaglio-Parrinello-SI}. We ran $50$ 
independent simulations, considering initial configurations either in the basin at $Z/Z_{\textrm{box}}\approx 0.36$ 
(escape rate), or at $Z/Z_{\textrm{box}}\approx 0.27$ (capture rate). 
We stopped the simulation when the system crossed the transition state region ($Z/Z_{\textrm{box}} \in [0.31,0.33]$).
To assess the reliability of the choice of the CV for the US simulations, we checked that no bias potential 
was added to the transition state region during the WT-metaD simulations~\cite{2014-JCTC-Salvalaglio-Parrinello-SI}.          
We also performed statistical analysis of the distribution of transition times. 
We performed a two-sample Kolmogorov-Smirnov (KS) test, which does not require a priori knowledge 
of the underlying distribution~\cite{2014-JCTC-Salvalaglio-Parrinello-SI}. 
We tested the null hypothesis that the sample of transition times extracted from the metaD simulations and a large 
sample of times randomly generated according to the theoretical exponential distribution reflect the same 
underlying distribution. The null hypothesis is conventionally rejected if the $p$-value $< 0.05$. The KS test has been 
performed as implemented in the software cran-R~\cite{cranR-SI}. The results are reported in the main text.

\end{document}